\documentclass[12pt,preprint]{aastex}

\slugcomment{Accepted for publication in ApJ, 10/19/2007}

\shorttitle{MID-IR FROM  HIGH MASS PROTOSTELLAR CANDIDATES}
\shortauthors{Campbell et al.}

\begin{document}


\title{MID-INFRARED PHOTOMETRY AND SPECTRA OF  THREE 
 HIGH MASS PROTOSTELLAR CANDIDATES
AT IRAS 18151-1208 AND IRAS 20343+4129 }


\author{M. F. Campbell\altaffilmark{1,12} , T. K. Sridharan\altaffilmark{2,12} , H. Beuther\altaffilmark{3} ,
J. H. Lacy\altaffilmark{4}, J. L. Hora\altaffilmark{5} ,Q. Zhu\altaffilmark{6}, M. Kassis\altaffilmark{7}, M. Saito\altaffilmark{8},
J. M. De Buizer\altaffilmark{9},
S. H. Fung\altaffilmark{10},  and L. C. Johnson\altaffilmark{11}}


\altaffiltext{1}{Department of Physics and Astronomy, Colby College, Waterville, ME 04901; mfcampbe@colby.edu }
\altaffiltext{2}{Center for Astrophysics,
    60 Garden Street, Cambridge, MA 02138;  tksridha@cfa.harvard.edu}
\altaffiltext{3}{Max-Planck-Institut f\"ur Astronomie, K\"onigstuhl 17, Heidelberg, Germany D-69117;
beuther@mpia-hd.mpg.de}
\altaffiltext{4}{Department of Astronomy, University of Texas,  Austin, TX 78712; lacy@astro.as.utexas.edu}
\altaffiltext{5}{Center for Astrophysics,
    60 Garden Street, Cambridge, MA 02138; jhora@cfa.harvard.edu}
\altaffiltext{6}{Center for Imaging Science, Rochester Institute of Technology, Rochester, NY 14623;
 qxzpci@cis.rit.edu}
\altaffiltext{7}{W. H. Keck Observatory, 65-1120 Mamalahoa Highway, Kamuela, HI 96743; 
mkassis@keck.hawaii.edu}
\altaffiltext{8}{National Astronomical Observatory of Japan, Osawa 2-21-1, Mitaka, Tokyo 181-8588, Japan; 
Masao.Saito@nao.ac.jp}
\altaffiltext{9}{Gemini Observatory, Casilla 603,  La Serena, Chile; jdebuizer@gemini.edu}
\altaffiltext{10}{Department of Physics and Astronomy, Colby College, Waterville, ME 04901, 
Present address: Department of Physics, 
Chinese University of Hong Kong, Shatin, NT, Hong Kong SAR, Peoples Republic of China; shfung@phy.cuhk.edu.hk}
\altaffiltext{11} {Department of Physics and Astronomy, Colby College, Waterville, ME 04901. 
Present address: Department of Physics and Astronomy, University of Wyoming, Laramie, WY 82071;
 lent.c.johnson@gmail.com}
\altaffiltext{12}{Visiting Astronomer at the Infrared Telescope Facility, which is operated by the University of Hawaii under Cooperative Agreement no. NCC 5-538 with the National Aeronautics and Space Administration, Science Mission Directorate, Planetary Astronomy Program.}


\begin{abstract}
We present arcsecond-scale mid-ir photometry (in the 10.5 $\mu$m N band and at 24.8 $\mu$m), and low resolution 
spectra in the N band ($R\simeq100$) of a candidate high mass protostellar object (HMPO) in IRAS 18151-1208 and
of two HMPO candidates in IRAS 20343+4129, IRS 1 and IRS 3. 
In addition we present high resolution mid-ir spectra 
($R\simeq80000$)
of the two HMPO candidates in IRAS 20343+4129.  These data are fitted with simple models to estimate the masses of gas and dust  associated with the mid-ir emitting clumps, the column densities of overlying
 absorbing dust and gas, the 
luminosities of the HMPO candidates, and the likely spectral type of the HMPO candidate for which [Ne II] 12.8 $\mu$m\ emission was detected (IRAS 20343+4129 IRS 3).  We suggest that IRAS 18151-1208 is a
pre-ultracompact HII region HMPO, IRAS 20343+4129 IRS 1 is an 
embedded young stellar object with the luminosity of a B3 star, and 
IRAS 20343+4129 IRS 3 is a B2 ZAMS star that has formed 
an ultracompact HII region and disrupted its natal  envelope.
\end{abstract}





\section{Introduction}

Many open questions in high-mass star formation are related to the evolution of 
circumstellar envelopes, accretion disks,  and jets from high-mass protostellar objects (HMPOs).  
HMPOs are often bright sources in mid-ir continuum, but only in a few recent cases have images suggested
specific structures such as disks, jets, or warm outflow cavity walls \citep{sri05, deb05, deb06,deb07}.   Mid-ir ionic lines like [Ne II] and [S IV] have
been used to map  compact 
and ultracompact HII regions (UC HII) and photodissociation regions, and to study their structure
and excitation \citep{lac82,oka01,zhu05,kas02,kas06}, but there is a lack of observations of HMPOs.
A remarkable feature of high-mass star formation is that HMPOs, defined as actively accreting mass,
can begin nuclear fusion and hence also be rapidly evolving massive young stellar objects (MYSOs) that have already formed hypercompact or ultracompact HII regions \citep{beu07}.  This feature raises a possibility of determining the spectral 
type of an MYSO through the ionic lines excitation, or from the number of
ionizing photons required for its observed centimeter continuum flux, separately from
estimating its luminosity and spectral type from infrared emission.  
However there is also the 
possibility that the  ionization is collisionally excited by a jet.  It may be possible to distinguish 
between the two cases, depending on the Doppler velocities, the morphology of ionized gas, and the 
ratio of the required flux of ionizing photons to total luminosity.  

Hoping to enlarge the sample of HMPOs that could be studied in detail despite potential
limitations, in 2003 we made mid-ir observations on the IRTF of about a third of the survey of 69 HMPO
candidates presented by \citet{sri02} and \citet{beu02a}. We chose objects from their survey that 
appeared compact  and/or bright 
in the MSX survey, and found that about 80\% of them were unresolved or marginally resolved by MIRSI \citep{deu02} on the IRTF in the broad N band at 10.5 $\mu$m and in a narrow band filter at 24.8 $\mu$m.  In addition, we obtained MIRSI grism low-resolution  spectra (R $\simeq100$) of ten of them in the N band.  In 2006 on Gemini North, we obtained TEXES \citep{lac02} high-resolution spectra (R $\simeq80000$)  of two HMPO candidates for which we had grism spectra.   In this paper we present spectra and photometry of three candidate HMPOs including the two with TEXES spectra: IRAS 18151-1208, IRAS 20343+4129 IRS 1, and IRAS 20343+4129 IRS 3 (hereafter, 18151, 20343 IRS 1,
and 20343 IRS 3).  
We will demonstrate that mid-ir emission from the dust  and gas near the HMPO candidates (where it is strongly heated) can be used as a useful probe of temperatures, masses, and luminosities, using simple isothermal clump models, even if each component (envelope, disk, jet, or cavity wall)  is
 not resolved.   In combination with observations cited below,
  we are able to use our new data to infer the nature 
 of each candidate candidate HMPO (e.g. pre-UCHII region HMPO, ZAMS B2 star).

The objects chosen are near the centers of complex, large-scale massive molecular outflows mapped by \citet{beu02b}.  20343 has an apparent large-scale N-S outflow whose red and blue lobes are both
extended E-W \citep{beu02b}, but 
 IRS 1 also has a compact E-W velocity outflow in CO(2-1), while  IRS 3 presents an ambiguous
 situation \citep{pal06}.
All objects show near-ir emission from shocked H$_2$ \citep{dav04,kum02}.  Two of them (18151 and 20343 IRS 3) were observed to have 0.5 and 1.8 mJy 3.6 cm emission, respectively \citep{car99,sri02}.  We observed 20343 IRS 1 and IRS 3  with TEXES on Gemini North based on the 3.6 cm and H$_2$ observations, with the goal of studying the role of ionized gas in them.

	The 10 $\mu$m grism spectral shapes of the HMPO candidates fall into three classes: those with deep silicate absorption; those with moderate silicate absorption and an apparent peak at about 8.5 $\mu$m; and those without an apparent silicate absorption feature but with continuum rising monotonically from short to long wavelengths (Campbell et al. 2007, in preparation).    Examples of these shapes can be seen in the UC HII spectra presented by \citet{fai98}.  Since the HMPO candidates
 were chosen based on IRAS colors similar to UC HII regions \citep{sri02}, one would expect the
HMPO candidates to have similar 10 $\mu$m spectra.  
IRAS 20343+4129, was observed with the IRAS LRS and has a silicate absorption feature \citep{vol91}.  The three objects presented here include an example of each of the three classes
of grism spectral shapes. 
Two of the ten candidate HMPO grism spectra show strong [Ne II] lines, IRAS 18247-1147 and 18530+0215.  \citet{sri02} reported relatively strong 3.6 cm fluxes of 47 and 311 
mJy, respectively, for them.  We did not observe them with TEXES on Gemini in order to focus on the
 earliest possible HMPO stage associated with ionized gas.

	Deriving physical information from the continuum spectra 
is difficult because the actual geometry of the dust distribution is unknown, except that the sizes of the N band and 24.8 $\mu$m images limit the extent of the emitting structures.  Experience with one-dimensional radiative transfer models of spectral energy 
distributions from UC HII regions \citep{cam95,cam00,cam04}, and inspection of the new spectra themselves suggest that there are ranges of temperatures in the emitting regions.  However, the two-dimensional radiative transfer models of \citet{deb05b} and \citet{whi03a,whi03b,whi04} show that 
 orientation of flattened envelopes and outflow cavities dramatically affects the depth of 
the silicate absorption feature, as does emission and absorption by 
 individual clumps in a clumpy dust cloud in the three-dimensional models of  \citet{ind06}.
The recent observations of extended and complex
near- and mid-ir emission from HMPOs \citep{deb05,deb06,sri05} also indicate that one-dimensional models are unrealistic.  Nevertheless, a simple three part model will allow us to derive approximate 
temperatures, column densities, and masses of different dust components. The first component
is hot dust that could be (part of)  a relatively compact disk, or a clump very near the HMPO candidate;
the second is warm dust that could be a more extended (part of a) disk, a clump of dust further out, or 
perhaps  the inner wall of 
of an outflow cavity; and the third is cold dust in an outer envelope that creates 
the silicate absorption feature.  

Deriving information from the ionic lines of an UC HII region is in principle straight-forward.  The line fluxes can
be   corrected for local extinction based on the continuum models discussed above, and
then the ratio of [Ne II] to [S IV] fluxes can be used to determine the exciting star's temperature
\citep{lac82,oka01}.  The numerical simulation code cloudy (http://www.nublado.org/)
can also be used to deduce the
star's temperature and luminosity 
by matching the simulated intensity and spatial extent of free-free, [Ne II], and 
[S IV] emission to the observations.
The star's temperature can then be compared to that of the
spectral type deduced from the cm continuum flux.   
In addition, comparison of Doppler velocities of the ionic lines to those
of molecular lines can be used to indicate if the gas is in an UC HII region or a jet.

\section{Observations and Reduction}
\subsection{MIRSI Images on the IRTF} \label{mirsi}

MIRSI, the Mid-InfraRed Spectrometer and Imager \citep{deu02}, was operated remotely from
Colby College and The Center for Astrophysics on the IRTF for these observations. 
MIRSI has an imaging field of view of
 $85 \times 65 \arcsec$ 
 on the IRTF, with diffraction limited performance ($0.8 \arcsec$ at $10 \mu$m),
and plate scales of $0\farcs2689$/pixel in right ascension and $0\farcs2635$/pixel in declination.  
The objects presented here were imaged on 2003 September 13.  From a trial
observation of HMPO candidates during a 2002 engineering run, we expected the sources to be unresolved or marginally resolved.  Conventional chopping and nodding was used with chopper throw $25 \arcsec$ NS and telescope nod $25 \arcsec$ EW, so that all chop-nod images  were on the array.  Five dithered
 integrations with total on-source time of 240 seconds
 were taken through a 5 $\mu$m wide N band filter centered at 
10.5 $\mu$m and then through a 7.9\% wide filter centered at 24.8 $\mu$m of the 18151 and 20343 fields.  Both  20343 IRS1 
and IRS3 were present in the latter field.  The dither pattern was a central position with $\pm 5
\arcsec$ offsets in right ascension and declination. 

Excess non-statistical electronic noise was removed with
a custom IDL procedure \citep{kas04}.\footnote{The MIRSI camera is divided into sixteen, 20-column channels, with each column containing 240 pixels.  For each image frame, the pixels in
all channels displayed a
distinctive noise pattern common to all that was determined and removed.   Each channel had a different median offset that was also removed. } Chop-nod addition and subtraction and flat
fielding were done in the usual way.  Negative chop-nod images were inverted, and all individual images from the dithered chop-nod sets were combined in IRAF.  

Positions were determined relative to 
those of calibration stars without special care for precision astrometry, so the positions are limited by the inherent offsetting accuracy and stability of the IRTF on that night to several arcseconds.  
Since 20343 IRS 1 and IRS 3 were observed simultaneously in each individual frame,
their relative positions should be accurate to better than two pixels ($0\farcs5$).
The relative positions of the mid-ir sources 
at the two wavelengths agree to $0\farcs1$ and they agree with the K band sources  of 
\citet{kum02} to $0\farcs5$, although it is not known if the mid-ir centers fall exactly on the K band
centers.  We assume that 18151 is coincident with the peak of 3.6 cm emission \citep{car99}.   Positions are shown in Table 1 for the sources.  

Simple photometry was performed using the IRAF task imexamine. 
The K stars $\gamma$ Aql and $\gamma$  Dra  were used for point spread functions (PSFs) and photometric flux density calibrations.  N band magnitudes for calibration were taken from the list of bright infrared standard stars on the IRTF web site.  We used the N band effective wavelength 
(10.47 $\mu$m) and the flux density for zero magnitude as given by \citet{tok00}.  For the MIRSI 24.8 $\mu$m filter, we used a zero magnitude flux density calculated by the formula given by \citet{eng92} shifted to agree with the N band zero magnitude
 flux density given by \citet{tok00}.  We assumed that the 24.8 $\mu$m magnitude
  would equal the  magnitude
in the  20.13 $\mu$m Q band \citep{coh95}.  This procedure gave agreement with the IRAS color corrected 25 $\mu$m flux density of 27.0 Jy for $\gamma$ Dra within 1\%.  We did not determine atmospheric extinction coefficients, and used the mean values for N and Q given by \citet{kri87}.     
N band flux densities appear to be reproducible to $\pm 10\%$ and to have a similar systematic accuracy, but the 
24.8 $\mu$m reproducibility and systematic accuracy is probably $\pm30\%$ due to the
variations in the atmospheric extinction over the course of the night.  The flux densities are shown in
Table 1, including the values used for  $\gamma$ Aql. 

Images of 18151 and 20343 IRS 1 appear to be unresolved at both wavelengths
 with indications of the first diffraction ring.  20343 IRS 3 is marginally resolved at both 
 wavelengths with FWHM a few tens of per cent larger than the PSF.  In addition, there is extended
 diffuse emission surrounding the central peak, especially in N band (Fig. 1a). The extended nature of the central peak of IRS 3 is clearer at 24.8 $\mu$m, and can be seen in 
 Fig. 1b.  
 FWHM values of Gaussian fits to the profiles by IRAF are shown in Table 1.   Our survey
 observations and ``pipeline''   image processing did not emphasize the 
 highest possible S/N for source and PSF profiles intended for image deconvolution.  We characterize the PSFs by values shown in Table 1 that are averages of 
Gaussian FWHM values and their standard deviations from calibration star observations
over the course of the night. 
The standard deviation in the PSF could be due to small errors in shifting
individual images before averaging and/or varying seeing in the unstable atmosphere.

\subsection{MIRSI Grism Spectra on the IRTF} \label{grism}

Grism spectra of the sources were obtained on 2003 September 14. The MIRSI grism was used 
with an N band prefilter, cutting off the spectra just longward of [Ne II] 12.8 $\mu$m.  The slit was 
oriented NS, and the system can be thought of as a long slit spectrometer.
On E. Tollestrup's suggestion, we arranged both chop and nod to be NS so that 
four spectra were recorded in each nominal 30 second integration camera frame,
giving total on-source time of 120 seconds per camera frame.\footnote{MIRSI's electronics adds multiple
chip frames, in this case of 700 ms duration,  recorded as extensions in fits files.  Separate fits
extensions are
recorded for each chop and nod position, and are then appropriately added or subtracted
to make up a camera frame. The nominal integration time would be the on-source time if 
the chopping and nodding were off the chip.}   We experienced considerable difficulty centering the source on the slit, and
only one of the chop-nod images could be well centered on the slit at best.  For each source,
multiple nominal 30 second integration frames were taken, although not all could be used
due to the source drifting off the slit or excessive noise.

The initial spectral processing involved the following steps: (1) For each chop-nod subtracted
camera frame, the electronic pattern noise was determined
from the  channel of the chip blanked by the N band filter's long wavelength cutoff, and
subtracted.  (2) The frames were flat-field corrected using a dome flat
from which a dark frame had been subtracted.  (3)
  All camera frames were averaged into a single combined frame
with the IRAF task combine.  (4)  Copies of the combined
frame were inverted and shifted as necessary, and then all four chop-nod spectra were averaged
to a single combined spectrum using the combine task.  
(5)  A bad pixel mask was applied to the single combined spectrum.  (6) Each  
channel's dark sky baseline surrounding the combined spectrum was inspected for an offset.  If an offset was found, the median level was shifted to obtain zero baseline outside the spectrum.  This processing proved
capable of recovering spectra where there was no signal visible in raw data frames.  18151
had strong signal visible, but both sources in 20343 appeared only barely visible after the
first stage of pattern noise correction.  IRS1 suffered from a noisy portion in its two best
frames between 10 and 11 $\mu$m.  22\% of the pixels in the spectrum in this wavelength band had excessive noise and were zeroed.  The final flux densities were corrected
for the zeroed pixels.

For the sources here, $\mu$ Cep, an M supergiant,  was used to calibrate the spectra. M supergiants are variable, and less desirable than the K giants used as photometric standards, but $\mu$ Cep 
gave an extremely high S/N calibration spectrum.  The $\mu$ Cep grism spectra were processed
in the same way as the candidate HMPO spectra.  
The IRAF task apall was used to extract the candidate HMPO spectra and the $\mu$ Cep spectrum.  
A linear grism dispersion relation was created using the telluric lines near 10 $\mu$m and the
[Ne II] 12.8 $\mu$m \ line.
The atmospheric
 transmission spectrum was taken from the UKIRT web site, convolved with the grism response, 
and cross-correlated to the uncalibrated MIRSI $\mu$ Cep spectrum
to determine one wavelength-pixel point.  The [Ne II] 12.8 $\mu$m line was used from the IRAS 18247-1147 spectum in which it was detected  for the second point.

In order to use our $\mu$ Cep spectrum for flux calibration,
we obtained ISO SWS spectra from the University of Calgary web site.\footnote{ http://www.iras.ucalgary.ca/satellites/$\sim$volk/getswsspec\_plot.html.}
This site uses a program to extract spectra as given by \citet{slo03}.  Three spectra are available for 
$\mu$ Cep, and they show some minor differences.  We chose TDT 05602852 for $\mu$\ Cep because it appeared best based on inspection of separate up and down scans kindly provided by \citet{kra06}.
A custom IDL procedure was used to calibrate the candidate HMPO 
spectra to the ISO  $\mu$ Cep flux densities. 
It multiplied each uncalibrated candidate HMPO spectral point by the ratio of the the ISO calibrated spectral point  (convolved to the same resolution as the MIRSI grism) divided by the uncalibrated MIRSI point for $\mu$ Cep.
This process gave us a preliminary calibrated spectrum.  
However, we did not know how well the sources were centered on the slit, so a final step was made,  
following a suggestion by E. Tollestrup.  For each preliminary calibrated spectrum, the flux densities, 
$F_\lambda$,  were summed over the spectrum, and the sum compared to the flux, $F$, as 
measured photometrically through the N band filter.  Their ratio was used as a correction factor for preliminary calibrated spectrum.  Flux densities in Janskys, $F_\nu$, were calculated from $F_\lambda$
for presentation in this paper.  The calibrated spectra are shown in Figure 2.  The 24.8 $\mu$m 
flux densities in Table 1 compared to the spectra indicate that the SEDs all rise significantly with 
increasing wavelength.

\subsection{TEXES Spectra on the Gemini North} \label{texes}
High spectral resolution observations of  20343 IRS 1 and IRS 3
were made during the
Texas Echelon Cross Echelle (TEXES) Demonstration Science  run on the
Gemini North 8-m telescope in July 2006 as part of the program GN-2006A-DS-2.
TEXES is a cross-dispersed mid-infrared (5-25 $\mu$m) spectrograph
capable of 0$\farcs$4 and 
3  km s$^{-1}$ resolution on Gemini
\citep{lac02}.  All data from the Demonstration Science run
are publicly available at http://archive.gemini.edu.

The TEXES candidate HMPO observations were made in the TEXES hi-med spectroscopic
mode with a 0$\farcs$5 slit giving $\sim$ 4 km s$^{-1}$ resolution.
The slit length was $\sim 4\arcsec$, oriented EW, with 0$\farcs$15/pixel
sampling along the slit.
Two observing modes were used: nod mode, in which the telescope was
nodded at $\sim$ 0.1 Hz to move the source by  $1\farcs7$ along
the TEXES entrance slit, and scan mode, in which the telescope was
moved south across the sky in 0$\farcs$25 steps without nodding to map the object.

Two spectral regions were observed, centered at [Ne II] (12.8 $\mu$m)
and at [S IV] (10.5 $\mu$m).
The spectral coverage at each setting was $\sim$ 0.6\%.
Spectral calibration was determined from sky emission lines, and
is accurate to 1 km s$^{-1}$.
Intensity calibration was obtained from observations of an ambient
temperature blackbody, and for scan mode observations is accurate to $\sim$ 20\%.
In nod mode,
there is an additional uncertainty due to the unknown fraction of the flux
outside of the  0$\farcs$5 slit.  For compact sources this is a random uncertainty due
to seeing and guiding, whereas for extended sources it causes a systematic
underestimate of the flux.
In addition, for both nodded and mapping observations the sky subtraction
procedure introduces an uncertainty due to the possibility of emission in
the sky position.

20343 IRS 1 was observed only in the nod mode at the [Ne II] setting.
Its 12.8 $\mu$m continuum  was clearly detected, with 11.5  Jy measured
through the 0$\farcs$5 slit.
This flux is in agreement with the MIRSI grism spectrum.
No scan map was made, but the source's extent
along the slit was less than 0$\farcs$5.
There was no evidence of the [Ne II] line in the spectrum, which is shown in Figure 3.
The equivalent width for a narrow ($<$ 10 km s$^{-1}$) line
was less than $2.0 \times 10^{-3}$ cm$^{-1}$,
indicating that the line flux from a 0$\farcs$5 region was less than
$6 \times 10^{-18} $ W m$^{-2}$  (2$\sigma$ uncertainties).
We note that the line flux could be greater than our limit if the line
source is more extended than the continuum source or if the line is broad.
A 100 km s$^{-1}$ wide line would have to be $2 \times 10^{-17} $ W m$^{-2}$ 
to be detected.
In addition, the line would have been missed entirely if it were at
V$_{LSR}$ = -20 to -50 km s$^{-1}$, which fell between TEXES grating orders,
but this would require a velocity shift of $>$30 km s$^{-1}$ from the
molecular cloud velocity of +11 km s$^{-1}$.

20343 IRS 3 was observed at the [Ne II] setting in both nod mode
and scan mode.  The nod mode observations have higher signal to noise
ratio, but have increased flux calibration uncertainty.
The nod mode spectrum and the scan-mode continuum and line maps are
shown in Figures 3, 4, and 5.
The 12.8 $\mu$m continuum flux, derived from a sum over a
$1\farcs35\times1\farcs25$ region where flux is apparent in the map,
is 2.3 Jy.  This is about 40\% of the flux in the MIRSI grism spectrum,
suggesting that some extended emission was missed.
For the nod mode observations, the nod throw was $1\farcs7$ EW, and for
the scans the sky background were taken from positions $1\farcs5$ north and south of
the peak. Consequently, emission on a scale $> 1\farcs5$ would have been missed.
The continuum source appears extended NS in the TEXES map, and possibly double-peaked.
Its extent is roughly consistent with the size of the MIRSI  10.5 $\mu$m N band image (Table 1;  \S3.1).
The more extended 24.8 $\mu$m emission is  elongated along a NW-SE axis rather
than NS (Figure  1; Table 1;
\S3.1). 
[Ne II] was clearly detected with a spectrally resolved line-to-continuum
ratio of $\sim$ 10.  The [Ne II] distribution appeared more point-like
than the continuum distribution, and  is located on the northern end of the 
extended continuum source (see Figures 4 and 5).
The line is centered at $V_{LSR} =15.7\pm1$  km s$^{-1}$ with an
observed FWHM $\sim$ 8 km s$^{-1}$.    
The line flux from the map is $1.0 \times 10^{-16}$ W m$^{-2}$. 

We attempted to observe IRS 3 in the [S IV] setting, but failed to detect
either continuum or line emission.
The MIRSI grism spectrum indicates that the 10.5 $\mu$m continuum is
a factor of about 2 weaker than the 12.8 $\mu$m continuum, so it should
have been detected, although not easily since the TEXES sensitivity is
a factor of about 2 poorer at 10.5 $\mu$m.
Given the possiblity of a pointing error, especially since this was
the first science run for TEXES on Gemini, we choose not to quote
an upper limit from these observations.

\section{Models of the Mid-IR Emission}

\subsection{Overview of a Simple Dust Continuum Model} \label{model}

The geometry of the mid-ir emitting dusty clouds is unknown except
for constraints on their projected diameters.  We can make simple three-component models to match
 our spectra and photometric data, and the models will give estimates of the 
 temperatures of the dust clumps,
 the column densities through them, the clump masses, and the mid-ir luminosities.  Our models
 should \it not \rm match the SEDs outside the mid-ir: we would expect them to underestimate the observed SED at both ends of the spectrum.  
 
 The model components are (1) a hot component whose size is constrained by the N band image, (2) a warm component whose size is constrained by the 24.8 
$\mu$m image, and (3) a cold, pure extinction component that is responsible for the silicate feature.  
In order to understand clearly the relationships of the various model inputs, we review the emission
of an isothermal, constant density, dusty clump observed through a colder cloud that creates 
extinction but no emission.  The observed flux density, $F_\lambda$, is given by
\begin{equation}
F_\lambda = \Omega B_{\lambda}(T) (1-e^ { -{\tau_{e}(\lambda) } } )  e^{- \tau_{a}(\lambda)}
\end{equation}
 where $\Omega$ is the solid angle of the clump, $B_{\lambda}(T)$ is the blackbody 
 intensity, $\tau_{e}(\lambda)$ is the optical depth of the emitting clump, and 
 $\tau_{a}(\lambda) $ is the extinction optical depth of the absorbing (and scattering) overlying
 cloud.  In general, emissivity, $\epsilon(\lambda)=(1-e^{-\tau_{e}(\lambda) } ) $, but
 for an optically thin clump
 $\epsilon(\lambda)=\tau_{e}(\lambda)$.  The optical depth in emission is given by
 \begin{equation}
 \tau_{e}(\lambda)=K_{e}(\lambda) \int \limits_{Clump} \rho_{d}dl
 \end{equation}
 where $K_{e}(\lambda)$ is the absorption cross section per mass of the emitting dust
 in cm$^2$ g$^{-1}$, 
 $\rho_d$ is the mass density of the dust, and the integral is through the emitting clump.
 The integral through the clump can be related to the column density of H nucleons, 
 $N_H$ through
 \begin{equation}
 \int\limits_{Clump} \rho_{d} dl = {{M_d}\over{M_g}} \mu m_{H}N_H
 \end{equation}
 where ${M_d}\over{M_g} $ is the dust to gas mass ratio, $\mu$ is the mean molecular
 weight for the assumed abundance ratios,  assuming neutral atomic gas instead of  H$_2$
 in order to follow the convention  of \citet{dra03a} that uses the 
 column density of H nucleons, 
 and $m_H$ is the mass of a hydrogen atom.
 Thus the emission optical depth can be related to the column density:
 \begin{equation}
 \tau_{e}(\lambda) = K_{e}(\lambda) {M_{d}\over M_{g}} \mu m_{H} N_H
 \end{equation}
 We can now clearly see from equation (1) that the observed flux density, $F_\lambda$ is
 related to both the solid angle and the column density, and that the derived column 
 density is strongly dependent on the assumed solid angle.  For the simple
 geometry of a constant density, end-on cylindrical cloud whose solid angle is estimated
 from the observed angle, $\theta_{FWHM}$, $\Omega={\pi\over4}\theta_{FWHM}^2$, and 
 in the optically thin case
 \begin{equation}
 F_{\lambda}={\pi \over 4}\theta_{FWHM}^{2}B_{\lambda}(T) K_{e}(\lambda){M_{d}\over M_{g}} \mu m_{H} N_H e^{-\tau_{a}(\lambda)}
 \end{equation}
 The assumed projected diameter of the clump is thus clearly a key factor in deriving a reasonable
 column density.  For optically thick clouds, the exponential function in the expression for 
 emissivity  makes the column density estimate
 very strongly dependent on the assumed diameter.
 
 A simple approximate way of estimating source diameters from the IRTF images for marginally resolved sources
 like 20343 IRS 3 (see Table 1) is based on assuming that the profiles of the
  true source, the observed data, and the PSF are all
 Gaussians.  With this assumption (that is certainly not correct for an Airy disk), a formal deconvolution would give
 \begin{equation}
 \theta_{s}=\sqrt{\theta_{d}^{2} - \theta_{p}^{2}}
 \end{equation}
 where $\theta_s$ is the true source FWHM, $\theta_d$ is the observed data FWHM,
 and $\theta_p$ is the PSF FWHM.  The resultant $\theta_{s}$ are $0\farcs75$ and $1\farcs57$
 in the N and 24.8 $\mu$m filters, respectively.  In this case however, the source extent at 
 the shorter wavelength 
 is more accurately determined by TEXES at 12.8 $\mu$m on Gemini. 
 Calculated in the same way, the extent
 is $0\farcs56 \times 1\farcs1$ EW $\times$
 NS, with area equivalent to a circular source of $0\farcs80$. 
 
 For the unresolved sources 18151 and  20343 IRS 1
 we can make estimates of the source sizes from the IRTF observations
  based on Gaussian deconvolutions as follows.
Let us represent $\theta_d$ in terms of $\theta_p$ and
 $n$ standard deviations of the PSF FWHM, $\sigma_p$: $\theta_{d}=\theta_{p}+n\sigma_{p}$.
 Substituting this expression into Eq(6) gives
 \begin{equation}
 \theta_{s}=\sqrt{2n\sigma_{p}\theta_{p}+n^{2}\sigma_{p}^{2}}
 \end{equation}
An observation of a source with a true value of $\theta_s$ in which the statistical variation
resulted in  a $-n\sigma_{p}$ deviation from $\theta_{d}=\theta_{p}+n\sigma_{p}$ would appear to be
unresolved with observed $\theta_d=\theta_p$.  Thus we can estimate 
reasonably likely values for source size $\theta_s$ from a $1\sigma$ deviation by substituting $n=1$
 into Eq (7),  and a $3\sigma$ upper
limit to a source size by substituting $n=3$.  
The resulting values of $\theta_s$ are larger than $\sigma_p$ or $3\sigma_p$ because of the deconvolution.  Table 2 presents diameters for the hot components and the warm components
calculated from the Gaussian FWHM values in Table 1 in N band and at 24.8 $\mu$m, respectively.  
They have been converted to AU at the distances of the HMPO candidates for the table.  The 
diameters in Table 1 are consistent with those of disk candidates associated with HMPOs cited by
\citet{ces07}, most of which have diameters from 1000-3000 AU.

 The derived column densities depend on the dust model that specifies 
 $K(\lambda)$.  A very attractive model is that of \citet{oss94} for protostellar cores
 often referred to as OH5.  This dust represents coagulated grains with ice mantles that
 would be expected to form from dust originally in the diffuse interstellar medium 
 during the process of molecular cloud formation.  It has been used to fit 
submm SEDs of high-mass protostellar cores \citep{van99,van00} and far-ir
observations of the UC HII region G34.3+0.2 \citep{cam04}.  However, when we examined
its $K(\lambda)$ behavior around 10 $\mu$m, we found that its silicate feature is shifted
longward from 9.7 $\mu$m and broadened so that it does not appear to be compatible
with our grism spectra.  Models by \citet{dra03a,dra03b} and colleagues fit the shape of our observed silicate absorption feature well.
We have chosen their 2003 $R_V=5.5$ synthetic extinction curve \citep{dra03a} for 
dust in dense clouds in the Milky Way. It is accessible  on the web and  
 well documented.  We have also fit the data successfully with $R_V=3.1$ dust 
 for diffuse clouds, but $R_V=5.5$ dust is appropriate for dense clouds and the deduced
 column densities should be more realistic. The dust model has 
 ${M_d\over M_g}=105$.  \citet{rom07} recently found the  $R_V=5.5$ dust
 to fit 1.2 - 8.0 $\mu$m data from the dense core Barnard 59.   In retrospect, we found high enough temperatures that ice mantles
 should have evaporated so that the OH5 dust would not be expected to fit our data.
 
 
 Fitting the shape of the grism spectrum of 18151 (Figure 2) and the large flux density 
 at 24.8 $\mu$m (Table 1) requires a minimum of three components for our models: hot dust responsible
 for the 8 $\mu$m end of the grism spectrum, warm dust for the 13 $\mu$m end of the 
 spectrum and the 24.8 $\mu$m
 photometry, and cold dust for the depth of the silicate absorption.  The model calculation has the
 following input parameters: the temperature of the hot component, $T_h$, the diameter
 of the hot component $\theta_h$, the optical depth in emission of the hot component 
 at 9.70 $\mu$m, $\tau_h$, the analogous parameters for the warm component, $T_w$, $\theta_w$, and 
 $\tau_w$, and the extinction optical depth, $\tau_a$,  of the overlying cloud that is too cold to emit in
 the mid-ir.  In the mid-ir, the extinction is virtually pure absorption.  The model uses 
 a spectrum that is the sum of the contributions of both hot and warm components
 each calculated according to Eq(1) to fit the observations.
 For given optical depths, column densities, $N_H$, can be derived from Eq(4)
 for each component, using parameters found in Draine's web site for the dust model.  
 For end-on cylindrical geometry, the masses of the hot and warm components
 can be derived from $N_H$ for each.  Visual extinctions are calculated from 
 the extinction cross section per H nucleon at the center of the V band as given in the 
 web site and $N_H$ for the cold absorbing component.  The SED due to the two dust clumps and 
 overlying extinction is  calculated to verify that it does not create excessive emission
 outside the mid-ir, and to derive the luminosity of the mid-ir emitting clumps for
 comparison to measurements and estimates of the overall candidate HMPO luminosity.
 
\subsection{Fitting the Continuum Data}\label{fitting}

Our goal is to apply our simple three-component model to the grism and 24.8 $\mu$m 
photometric data to derive estimates of the temperatures, 
the column densities, the masses,  and the luminosities of the emitting
dust components, and the column density of the cold extinction component.  We have
no measurements of the angular extents of the components for 18151 or
20343 IRS 1 because they were unresolved, although we do have them
for 20343 IRS 3.  If we
assume incorrectly small  sizes for the first two, we will make over-estimates of the 
column densities.  We have chosen to use large size estimates that are consistent
with the data for these two so that our column densities can be thought of as lower limits
to the true values. The parameters for our models are shown in Table 3.  Table 3 shows sizes
based on 1 $\sigma$ deviation estimates for  unresolved sources in Table 2.  For 
20343 IRS 3, it uses the size from the TEXES continuum map for the hot component,
and the size based on Gaussian deconvolution of the MIRSI 24.8 $\mu$m image for the warm
component.

The models were interactively fit to the data.  In order to quantify the quality 
of the fit and to aid in choice of extreme values of the parameters consistent with the systematic 
accuracy of the data, we defined a modified $\chi^2$ statistic.  The grism spectra were broken into four 
photometric bands: $8.0-9.0 \mu$m, $9.0-10.5 \mu$m, $10.5-12.0 \mu$m, and $12.0-13.0 \mu$m. 
Fluxes were summed in each band.  The 24.8 $\mu$m photometry served as a fifth band.  We defined
 the modified $\chi^2$ statistic for each band, $i$, as 
\begin{equation}
\chi^{2}_{m i}  =  \frac{  { (F_{i, Model}-F_{i,Data} ) }^{2} } { {F_{i,Data}}^2 } 
\end{equation}
 where
 $F_i$ is the flux in band $i$.   The sum of the
five terms formed $\chi^{2}_{m}$ for the model.  This statistic places the quality of the
fit as a fraction of the data value in each band on equal footing.  Even though it assumes 
 that each of the five bands has equal S/N and has no specific
statistical interpretation, it is useful for fitting models to the data.  
For optimizing models, we sought to minimize $\chi^{2}_{m}$ interactively, initially 
using graphs of the models plotted over the data as guides, rather than
using an exhaustive, automated search through the parameter space.   
Such an automated search is not
justified by the quality of the data and the simplicity of the model.  
Final models in Table 3 were optimized using $\chi^{2}_{m}$ values. 
For choosing the upper or
lower value of a
specific parameter (e.g. $T_w$) that might be reasonable, we started with the value
in an optimum model, and varied that parameter (only) until the band most 
affected by the parameter (in this case $i=5$, $24.8 \mu$m) indicated
a 30\% difference (our systematic photometric accuracy), or $\chi^{2}_{m i} = 0.09$.    On a graph
of the data like Figure 6, the upper value $T_w$ would cause the
model's flux density  to fall on the end of the upper error bar at 24.8 $\mu$m.  
It
was not feasible to vary more than one parameter at a time because of the size of the 
parameter space and the fact that all bands of our data were somewhat affected by all of the
parameters.  The procedure is arbitrary.  Larger extreme values for a parameter of interest 
could be found if all others were also varied compensating for the changed parameter of
interest (e.g. if $T_h$ were the parameter of interest, a larger extreme value would
result if $\tau_h, T_w, \tau_w,$ and $\tau_a$ were varied in addition to $T_h$).

The best-fit temperatures of the hot and warm components are 
in the ranges 420-1000 K and 110-182 K, respectively.  
It is useful to examine SEDs over a wide range of wavelengths in order to understand 
how the components combine to fit the data.  We show the SEDs of individual components in
Figure 6 for the model of 18151 with a 455K hot component whose parameters are
given in Table 3.  Figures 7 and 8 show models for 20343 IRS 1 and IRS 3, and their
model parameters are also shown in Table 3.
In addition, Figures 6-8 show the combined SEDs both with and without the cold 
components' extinction,  
 and the IRTF data.  In Figure 6, 
 the 455K hot component shows a strong 9.7 $\mu$m silicate emission feature, as
expected.  However the 136K warm component does not show a strong 
emission feature because its Planck spectrum 
is rapidly rising to long wavelengths and the silicate feature is smoothed out
because it is optically thick in its center.
Absorption features at 9.7 and 18 $\mu$m due to the cold layer are clear in the total spectrum, 
and the latter causes a nearly flat portion between 15 and 20 $\mu$m.  
This shape in the mid-ir is not necessarily an artifact of the simplicity of a
three component model.  A similarly shaped SED is shown from the library of 
Monte Carlo radiative transfer models at http://caravan.astro.wisc.edu/protostars/
for the low mass protostar IRAS 04368+2557 by \citet{rob07}.  Our model SED 
for 
18151 in Figure 6 also
clearly shows that these components emit very weakly at both near-ir and submm 
wavelengths.  For 18151, a large
range in $T_h$ will fit the data 
 apparently because the grism spectral range
could lie in the Raleigh-Jeans part of the hot component's spectrum for high values of
$T_h$.   The  $T_{h}=455$ K model is in the lower end of the range.
A model with $T_{h}=1600$ K, a commonly assumed dust sublimation
temperature \citep{whi04}, will fit our data, but would have a large excess over the observed
flux at 2.1 $\mu$m \citep{dav04}.  A model with  $T_{h}=1000$ K is presented in Table 3 that fits both
our data and the flux at 2.1 $\mu$m.

\subsection{General Results and Discussion for the Continuum Models}\label{models}

Parameters for the models are presented in Table 3.  Two models are given for 18151 
at extreme ends of the range of $T_h$ that fit the data well, and the best-fit model is given for each of 
20343 IRS 1 and IRS 3. The summed values of $\chi^{2}_{m}$ are given in the last 
column.\footnote{Of the models, $\chi^{2}_{m}$ was largest for 20343 IRS 3.  Its components,
$\chi^{2}_{m,i}$ are 0.0005, 0.0008, 0.0035, 0.0093 and $2.7 \times 10^{-6}$ for $i=1$ to 5,
respectively.}
 In addition, upper and lower extreme values that are
consistent with the data to within 30\% as discussed above are shown.

There are a number of important
 aspects of the values derived from the models.  The first is
that the source that appears most deeply embedded, 18151, could have a hot component 
with a sufficiently high temperature so that its mid-ir would be on the Raleigh-Jeans end of the spectrum,
so that the temperature cannot be determined without measurements in shorter bands like K,  L or M.
However, it is entirely likely that some or all of the flux at these shorter bands would come from an 
additional, hotter dust component than the ones whose mid-ir we seek to model.
For the other two objects, Table 3 shows moderate ranges in temperatures from lower to upper values 
that are compatible with the best-fit emission optical depths.  The range of temperatures that
could be fit by varying the emission optical depths at the same time with only modest increases in $\chi^2$ would be larger than shown.   Nevertheless, we feel the temperatures shown are 
realistic estimates.
The ranges in $T_h$ and $T_w$ are well separated in all of the sources.   If each source's emission is 
from a single clump or disk with a continuous range in temperatures,  each must contain a 
wide range in  the actual temperatures.

It is interesting to note that the temperatures for what we call warm dust, $T_w$, are rather close
to those of the ``hot component''  of \citet{sri02} based on IRAS data ($T_{hd}$, of 170 K and 150 K
for 18151 and 20343, respectively).  It suggests that our IRTF measurements are measuring
much of the same dust as the shorter IRAS bands measured. In fact, our N band $F_\nu$ 
values for 18151 and 20343 IRS 1 + IRS 3 are each slightly more than 50\%  of the IRAS
PSC values.  Our 24.8 $\mu$m $F_\nu$ for 18151 virtually equals the IRAS PSC 25 $\mu$m
value,  and the sum of the 20343 24.8 $\mu$m $F_\nu$ values is 2/3 of the IRAS 25 $\mu$m value.

Another interesting aspect is the extremely small amount of gas and dust in the hot components.  
This hot material is unlikely to  be a major part of 
accretion disks that might be expected for  HMPOs, because the accretion disks are 
likely to contain about the same mass or more as the stars, and to have
characteristic temperatures from ten to several hundred K \citep{ces07}.
 It could be in a hot inner rim or surface layer of a photoevaporating disk \citep{hol94}.  
The assumed diameters of the mid-ir components suggest that
the emission comes from material that might be described as
being in an outflow cavity wall.  In fact, it may be coming from the intersection of 
a flared accretion disk with the surface of the outflow cavity,  as has been suggested
for other sources by \citet{deb07b}.  Although the assumed diameters of
the hot components are somewhat arbitrary and affect $N_{H,h}$, they do not affect the mass
estimates since these components are optically thin.   The ranges about the central value of
$N_{H,h}$ and $M_h$ are about $\pm 30\%$, as expected for optically thin sources.  (Lower limits
to $N_{H,h}$ based on the 3$\sigma$ upper limits on source diameters shown in Table 2 can be 
calculated from the data in Table 3 since $N_H$ is inversely proportional to $\Omega$
for optically thin sources.)  Accurate estimates of  $N_{H,h}$ and 
number density $n_{H,h}$ of the hot components of 
 18151 and 20343 IRS 1 will require higher resolution observations on a larger telescope.
 
 The optical depths in emission of the warm components, $\tau_w$, are high in the cases
  of 18151 and 20343 IRS 3, 4.2 and 0.9, respectively.  This effect is a surprise because the assumed values of the projected diameters $D_{w}$
 are not particularly small compared to observed and expected diameters for structures 
like candidate accretion disks near HMPOs, 
 \citep{deb05,she01, ces07}.   While 18151 was unresolved and its value for $N_{H,w}$ is essentially
 a lower limit,   IRS 3 was resolved, so its  value for $N_{H,w}$ is a firmer
 estimate.  For these two HMPO candidates, indicated $N_{H,w}$ values 
 are about as large as the column density for extinction, $N_{H,a}$.  
  Even with these high values,
 however, the masses are less than 1 M$_{\sun}$ and much less than the mass expected for
 an HMPO.  This situation again indicates that the mid-ir emitting
  dust is not tracing the bulk of the mass expected to be in accretion disks.
 
 Our mass estimates have led to the conclusion that in these three HMPO candidates, mid-ir emission is
 not indicating massive accretion disks.  Either  only
 a small fraction of their mass is emitting in the mid-ir, or the disks have been disrupted already. 
 The emission may well come from
 dust in and around the  walls of outflow cavities as has been suggested
 for other HMPOs by \citet{deb05} and \citet{deb06,deb07} and as is indicated in the two dimensional
 radiative transfer models for a low mass class I protostar of \citet{whi03a}.  
  
 Extinction optical depths, $\tau_a$, are not extremely high, and cover a limited range.   There is
 a selection effect:  if they were larger, the HMPO candidates would not have been included in the original
 survey for lack of 12 $\mu$m IRAS detection, or we might not have detected the HMPO candidates
  in the mid-ir
 on the IRTF.  (Five of 23 fields chosen from the \citet{sri02} survey resulted in
 non-detections in N band on the IRTF.)
 The appearance of the 20343 IRS 3 grism spectrum suggested that there might not be
 any extinction, but fitting the spectrum required all of the components.  In some
 cases, the range in
 $\tau_a$ from lower to upper value is small because the extinction is exponential
 and the sources are optically thick in extinction.  The values of $N_{H,a}$
 and $A_V$ are also not extreme since they are directly proportional to $\tau_a$.
 Unlike the emitting components, the extinction $\tau_a$ and parameters derived from 
 it ($N_{H,a}$ and $A_{V,a}$) are not affected by the assumed source size, and hence are
 more firmly defined values.
 
 High mass stars are expected to form at the centers of cluster-forming molecular
 cloud cores \citep{beu07}.  In our small sample of three objects, only 18151 appears to
 be centered on a 1.2 mm continuum core in the plane of the sky.
   Comparison of  the column density of H nucleons per cm$^2$, $N_{H,a}$, and the
   visual extinction in magnitudes, $A_{V,a}$, from the mid-ir model to the values for column density of H$_2$, $N_{gas}$, and  $A_V$ from 1.2 mm observations given
 by \citet{beu02a} would indicate if the candidate HMPO is indeed centered along the line of sight within
 the larger 1.2 mm dust core
 if a consistent dust model and units were
 used at both wavelengths.  If the HMPO candidate were at the core center, interior to the
 bulk of the 1.2 mm dust core's mass, 
 we would expect the mid-ir based column density for the cold component to be due to near 
 side of the  
 dust core, and have  one-half the column density based on the 1.2 mm observation.
 For a consistent comparison, we have calculated $N_H$ from equation 
  (5) using \citet{dra03a} dust, the peak flux of 673 mJy for $(11\arcsec)^2$ at 1.2 mm given by \citet{beu02a} in their Table 2, 
  $T_{dust}=T_{cd}=47$K given by \citet{sri02} in their Table 1, and the solid angle of $(11\arcsec)^2$
  in place of ${\pi\over4}\theta^2_{FWHM}$.  
  The \citet{dra03a} $R_V=5.5$ dust model has
  $K_e(1.2 \rm{mm})=0.2388$  cm$^2$\  g$^{-1}$ and $A_V=7.29\times10^{-22}N_H$, where
  $N_H$ is the column density of H nucleons.  For 18151, the 1.2 mm results are
 $N_H=5.6\times10^{23}\ \rm cm^{-2}$ (log$N_H=23.75$) and 
  $A_V=410$.  One half of the column density would give log$N_H=23.45$ and 
  $A_V=205$.  These values are a factor of 2.8 larger than the mid-ir based values in Table 3, log$N_{H,a}=23.00$, and 
  $A_{V,a}=72.1$.    It appears that the mid-ir source is not at the center of the 1.2 mm emitting
  core, but near its front side. However, there are significant uncertainties in the estimates.  We have ignored the column
  densities of hot and warm dust in emission in the mid-infrared,  but their size $\sim1\arcsec$
  on the sky
  should not have contributed significantly to the 1.2 mm emission detected in an 11$\arcsec$
  beam.
  The largest uncertainty lies in the value of 
  $K_e(1.2 \rm{mm})=0.2388 \ cm^2\  g^{-1}$.  For far-ir through mm wavelengths, the
  emission coefficient is often approximated as $K\propto \lambda^{-\beta}$.  \citet{dra03a} 
  $R_V=5.5$ dust
  has $\beta=1.8$ between 250 $\mu$m and 1.2 mm.  
  Based on a sample of 69 HMPO candidates, \citet{wil04} derive a mean 
 $\beta$ of 0.9\footnote{For  18151, \citet{wil04}
  found $\beta=0.5$ for the limited range of 450-850 $\mu$m.  However, $\beta \simeq1$ is 
  much more commonly cited for HMPO candidates. }, that would increase 
  $K_e(1.2 \rm{mm})$ by 4.1, and decrease the column density
  to give log$N_H=22.84$ and $A_{V,a}=50.1$, in much better agreement with the mid-ir based
  results.  In fact, one could turn the problem around.  For cases where there is strong evidence
  that an HMPO is centered in a core, the ratio of column densities determined by the extinction
  at 9.7 $\mu$m and 1.2 mm dust emission determines the overall slope of the extinction curve between
  them, if one assumes a wavelength in the far-ir at which $K_e(\lambda)$ flattens.   Using
  250 $\mu$m as the reference wavelength (e.g. \citet{hil83}), the mid-ir derived
  column density  $N_H=1.0\times10^{23}$ cm$^{-2}$, and 1/2
  of the  1.2 mm flux density to account only for the emission in front of the HMPO,
  we find $\beta=1.2$.

%

  There are other possible reasons for discrepancies.  For cores that do not fill the 1.2 mm
  radio beam, one would expect the difference in 
  telescope beams to affect the ratio of column densities, with the larger beam 1.2 mm
   $N_H$  being less than the smaller infrared beam  $N_H$.  Conversely, 
  an unresolved clumpy structure in the 1.2 mm emitting core could result in higher 
  1.2 mm beam-averaged value than observed in a small diameter infrared beam that happens
  to pass between clumps giving a low extinction line of sight to the
  HMPO.  Similarly, viewing the source along an
  outflow cavity as modeled by \citet{whi03a,whi03b,whi04} 
  would result in  lower ir-based $N_H$ and $A_V$.  However, the outflows observed for
  these objects appear to lie in or near the plane of the sky (see \S 4).
  
  The HMPO candidates 20343 IRS 1 and IRS 3 are not projected against peaks in the 1.2 mm emission
  \citep{pal06}.  Nevertheless we can estimate $N_H$ and $A_V$ for them from the 
  1.2 mm  map  \citep{beu02a,pal06} that shows $\sim200$ mJy/($11\arcsec)^2$ for both.  The ratios of mm-based 
  $N_H/2$ to ir-based $N_H$ are 1.5 and 5.9 for IRS 1 and 3, respectively, in comparison to 
  2.8 for 18151.  These are consistent with IRS 1 being deeply embedded but not the apparently 
  more evolved IRS 3.
  
 
 Observations that cover both mid- and far-ir wavelengths are commonly used to estimate the total 
 luminosity of an embedded source (e.g. \citet{sri02}).  For a case in which an HMPO is
 completely embedded so that the observed infrared emission comes from a circumstellar dust
 envelope that absorbs and reradiates all of the HMPO's emission, the total luminosty is 
 accurately measured by the integrated SED.
 Far-ir observations like IRAS data cover the peak of the
 SED, but the large beams can include sources in addition to that of the 
 HMPO of interest, especially since high mass stars usually form in clusters.
   The higher resolution mid-ir observations can be
 limited to a single HMPO, but lack the far-ir contribution needed for
 total luminosity.  Consequently, luminosity  based solely on mid-ir fluxes is 
 usually quoted as a lower 
 limit to total luminosity.  Our simple models can be used for lower
 limit 
 luminosity estimates that account for extinction in the mid-ir and include
 some flux outside the mid-ir.  In Table 3,
 we show  the integrated model SED fluxes  that would be observed through the 
 extinction (the red curves in Figures 6-9) as $L_{o}\over L_{\sun}$.  We
 take these model based values as lower limits to the HMPO candidates' luminosities that are somewhat
 improved over simply adding our observed mid-ir flux densties. 

In Table 3, we also show the summed modeled luminosity emitted by the hot and warm clouds
together without the extinction factor applied (the orange curves) as $L_{e}\over  L_{\sun}$. 
If the hot and warm dust components had absorbed all of an HMPO's power and emitted 
(re-radiated) it as modeled, the sum of their SEDs would give
the total luminosity.   Presumably, the cold extinction layer re-radiates in the far-ir the 
energy it absorbed from the hot and warm components, so 
an observed full SED extending into the far-ir would
 have the same luminosity  as $L_{e}\over L_{\sun}$.  An outflow cavity
 should not have much effect on the the luminosity estimate.  The mid-ir portion
 of a source's SED is affected by the line of sight to the cavity, but the far-ir
 portion that dominates the luminosity is not as  strongly affected \citep{whi03b}.  
 Outflow cavities for our sources would have little effect because they appear to be
 close to the plane of the sky (see \S 4).   Our $L_{e}\over L_{\sun}$
 value should be close to the luminosity based on IRAS presented by
 \citet{sri02}, if we have observed the dominant HMPO 
and there is little luminosity from the other YSOs. 
 For 18151, our estimate is 22400 $L_\sun$ in close agreement  with 
 the IRAS-based value of 20000 $L_\sun$  \citep{sri02}.  
 The HMPO candidates 20343 IRS 1 and IRS 3 lie in a single IRAS beam.
 For 20343 IRS1 and IRS3 combined, our value of $L_{e}\over L_{\sun}$
  is 2200 $L_\sun$, about two-thirds
  of 3160 $L_\sun$ from IRAS.  The star in 20343  IRS 3 is apparently more evolved than the HMPO
  candidate in 
  18151 (see \S 4).   As a consequence, it  has a lower column density envelope ($N_a$) than 
  18151, so its envelope does not absorb the full stellar emission.  
  
  Overall, the $L_{e}\over L_{\sun}$ luminosity estimates  agree well
  with IRAS.  Earlier we noted remarkably close agreement between
 our flux densites and those of the IRAS PSC, and between our $T_w$ and the
 IRAS-based $T_{hd}$ of \citet{sri02}.  Consequently the agreement of luminosities is
 not surprising.

 \subsection{Analysis of High Resolution Spectra}\label{lines}

The infrared fine-structure lines are collisionally excited forbidden lines.
As such, they have emissivities proportional to the product of the
density of the emitting ion, the electron density, and the collisional
excitation cross section, $N_i N_e q_{lu}$, for $N_e << N_c$
\citep{ost06}.
$N_c$, the electron density at which the collisional deexcitation rate
equals the radiative deexcitation rate, is $5 \times 10^5$ cm$^{-3}$ for
[Ne\,II] and $2 \times 10^4$ cm$^{-3}$ for [S\,IV].
The collisional excitation rates are proportional to
$T^{-1/2} e^{-E_u/kT}$, or about $T^{-0.4}$ for these lines.
Since the radio free-free emissivity is proportional to
$N_e N_p T^{-0.35}$, the ratio of the fine-structure line fluxes to
the free-free continuum flux is proportional to the ionic abundance
relative to that of ionized hydrogen, with a weak dependence on electron
temperature and density.
The ionic abundances relative to the total atomic abundances depend most
strongly on the spectral type, or effective temperature, of the ionizing
star, with weaker dependences on the stellar luminosity and the electron
density.
Consequently, the ratio of the fine-structure line fluxes to the free-free
flux and to each other can be used to determine the stellar spectral type.

The most convenient way to determine the stellar parameters that are
consistent with the measured fluxes is with a nebular modeling program,
such as cloudy \citep{fer98}.
We used cloudy (version 07.02.00) to calculate several dust-free\footnote{
The presence of dust in an HII region affects both the free-free and
[Ne\,II] fluxes in the same way.  It could cause $N_{Lyc}$ for the star to
be underestimated.  Evidence that $N_{Lyc}$ were underestimated would come from
an ir-based luminosity larger than expected from $N_{Lyc}$ and stellar models.  That does not appear
to be the case for 20343 IRS 3 (see \S4,2).}
models with
stellar parameters similar to those of a B2 star, $T_{eff} \sim$ 20,000 K and
$N_{Lyc} \sim 4 \times 10^{44}$ photons s$^{-1}$.
An acceptable fit to the observed free-free and [Ne\,II] fluxes from
20343 IRS\,3 was found for $T_{eff}$ = 19,000 - 22,000 K and
$N_{Lyc} = 3-5 \times 10^{44}$ photons s$^{-1}$, assuming $N_e << 5 \times 10^5$ cm$^{-3}$.
The [S\,IV] flux was predicted to be much less than that of [Ne\,II],
consistent with our failure to detect [S\,IV].

The extinction-corrected luminosity of IRS\,1 is 1380 L$_{\sun}$ and
 neither radio free-free nor [Ne\,II] emission was detected
from it, with limits of about 1/3 and 1/15 of the free-free and [Ne\,II]
fluxes from IRS\,3.
The luminosity and lack of observable ionized gas are consistent with a spectral type
of B3 or later, or $T_{eff} <$ 18,000 K, as the ionizing flux of a B3
star is only about 1/10 that of a B2 star \citep{pan73}.

 \section{Discussion of Individual Sources}\label{sources}
  \subsection{IRAS 18151-1208}\label{I18151}
  This HMPO candidate was chosen for study because of its deep silicate absorption in the mid-ir (Figure 2), 
  the location of its mid-ir peak at a 1.2 mm peak in the \citet{beu02a} survey, its low level of 3.6 cm emission \citep{sri02,car99}, its large-scale CO outflow \citep{beu02b}, its detection in the
  K band, and the presence of H$_2$ molecular jets \citep{dav04}.   At  mm and sub-mm,
   there is a relatively
  simple peak 13$\farcs2$ east and $4\farcs9$ south of the IRAS position with some extension to the southwest  \citep{beu02a,wil04}.  There is a strong second 1.2 mm peak about $100 \arcsec$ west
  and 33$\arcsec$ south form the IRAS position, and there are two weaker peaks $>80\arcsec$ from  it
  \citep{beu02a}.  The main sub-mm peak is within $\sim2\arcsec$ from the MSX source that we identify
  with our mid-ir source \citep{wil04}.  
  The IRAS and mid-ir luminosity estimates agree well at $\sim 20000$ L$_\sun$ (\S 3.3), 
close to that of a B0 ZAMS star \citep{pan73}.  \citet{dav04}
  characterize the K band emission as showing a dense cluster.  They conclude that
  IRS 1, the source at the K band peak, is a pre-UC HII HMPO.   It  is the only 
  deeply embedded near-ir object in the field \citep{dav04}, and it is
  coincident with the mm peak and with an 0.5 mJy 3.6 cm source \citep{car99}.

 For an optically thin HII region, the 3.6 cm flux density of 0.5 mJy would require log $N_L$=44.67
  ionizing photons s$^{-1}$ at 3 kpc, using the common relationship 
  between flux density and number of ionizing
  photons as given by \citet{deb05s}.  This number is that of a
  B2 ZAMS star \citep{pan73}, that would have $L=2900$ L$_\sun$ compared to the ir-based luminosity 
  of $\sim 20000$ L$_\sun$.  The order of magnitude
  difference in luminosity
  could be explained by a significant contribution from lower luminosity members of the cluster,
  but this seems unlikely since the 24.8 $\mu$m diameter has a 3$\sigma$ upper limit 
  of 3670 AU (Table 2).  
 Apparently,  an HMPO  at IRS 1 at the pre-UC HII stage is creating the ionization
 as a  hypercompact HII region (HC HII) or as a jet.   (In either case the equation used for 
  $N_L$ does not apply.)  In
 fact, \citet{dav04} found two lines of clumps of shocked H$_2$ indicating the presence of two jets, 
 one of them centered on IRS 1, and detected Br$\gamma$ at IRS 1. Assuming $A_V=30$, they estimated an
accretion luminosity of $\sim 120 $ L$_\sun$ and inferred a total $L\sim 1000 $ L$_\sun$
 by extending results from low
luminosity YSOs.  They argued that these are lower luminosity limits because their $A_V$ was
calculated for the extended H$_2$ flow rather than for the source center where we have found
$A_V=72$.  Our value suggests increasing the extinction correction for Br$_{\gamma}$
by 83 using \citet{dra03a} dust, resulting in an inferred luminosity well in excess of 
20000 L$_\sun$.  

  The large scale CO outflow red and blue shifted lobes are complex \citep{beu02b}.  
 The two jets of H$_2$ clumps are nearly at right angles, and \citet{dav04}
 interpret the complex CO  outflow as consistent with them.  Both of the H$_2$
 jets are close to the plane of the sky \citep{dav04}, so it is unlikely that our 
 line of sight is along a low extinction outflow cavity. 
  The overall CO outflow distribution and the more powerful
   H$_2$ jet both have their centers near IRS 1, 
  while the second jet's center appears to be about 10$\arcsec$ SW of IRS 1, well outside the
  mid-ir emission.  \citet{dav04} estimate the luminosity of the jet centered on IRS 1 to be
  0.7  L$_\sun$, and the second jet to have only $\sim0.05$ L$_\sun$.  The H$_2$ jet 
  luminosity of   0.7  L$_\sun$ is
  much higher than those in low mass YSOs \citep{dav04}.  A larger
  extinction correction for the H$_2$ luminosity, as suggested above, will increase
  it significantly.   A large outflow luminosity would suggest that it is still strongly accreting.
  
  
  This source has a Class II CH$_3$OH maser at its mm peak, but neither an H$_2$O maser,
   \citep{beu02c}, nor an OH maser \citep{edr07}.  (An H$_2$O maser is associated with the second 1.2 mm source in the 18151 field.)   
 \citet{beu02c} summarize the conditions for models of radiative pumping of the
  CH$_3$OH masers as $T\sim150$ K,  methanol column density $N_M> 2\times10^{15}$ cm$^{-2}$,
  and $n_H<10^8$ cm$^{-3}$.
  These are are quite close to the parameters of the warm dust in our
  mid-ir emission model that has $T_w=136$K,  and $N_H=8.5\times10^{22}$ cm$^{-2}$ with a
  diameter of 2050 AU.  If the source has a line of sight dimension equal to its diameter,
  the density is $n_H=2.9\times10^6$ cm$^{-3}$.  A recent set of models for 
  CH$_3$OH maser cites fractional methanol abundance $X_M=10^{-5.7}$ 
  \citep{cra02} that would
  give methanol column density $N_M=1.7\times10^{17}$ cm$^{-2}$.  \citet{cra02} cite
  $N_M/\Delta V$ where $\Delta V$ is the linewidth as a key parameter.  They give 
  $\Delta V=1$ km s$^{-1}$ as typical.  With it, our model warm dust component 
  would have $N_M/\Delta V=10^{12.23}$ cm$^{-3}$ s.  Its conditions appear to be comfortably within
  the conditions for strong 6.7 GHz emission (\citet{cra02} Figure 1). Thus our mid-ir based
  model density and temperature are consistent with the observed methanol maser
  emission, and with a pre-UC HII stage for 18151.
  
  For OH masers, the use of $X_{OH}=10^{-6}$ \citep{cra02} gives
   $N_{OH}/\Delta V=10^{12.2} $cm$^{-3}$ s.
This value and $n_H=2.9\times10^6$ cm$^{-3}$ are in a region of parameter space where
OH masers are unlikely  for $T_k=150$ K (\citet{cra02} Figure 1), consistent with observations.  

	There have been tentative suggestions that masers
   might appear in an overlapping sequence of CH$_3$OH, H$_2$O, and OH in
HMPOs  \citep{beu02c, cra02}, but  the evidence does not seem conclusive \citep{deb05s}.
 If the sequence were correct it would suggest that 18151 is an early stage HMPO.   Overall, our mid-ir observations and models strongly support the proposition 
that 18151 IRS 1 is an  pre-UC HII HMPO, with a luminosity suggesting type B0.
 
 \subsection{IRAS 20343+4129 IRS 1 and IRS 3}\label{I20343}
 
 IRAS 20343+4129 is one of the brighter sources in the mid-ir in the \citet{sri02} list of
 HMPO candidates, and its IRAS LRS spectrum shows a clear silicate absorption feature at 9.7 $\mu$m
 \citep{vol91}.  At a relatively close distance of 1.4 kpc, its IRAS based luminosity is 
 only 3200 L$_{\sun}$ \citep{sri02}.  It has weak 3.6 cm continuum emission,
  but not  any maser emission \citep{car99,sri02}.
 \citet{beu02b} found it to have two massive molecular outflows.  The stronger 
 massive outflow
 is close to the near- and mid-ir sources, but its outflow luminosity 
  is among the weakest they observed. \citet{kum02} found three K band
 continuum YSOs with compact circumstellar H$_2$ emission.  Two of them,
 IRS 1 and IRS 3, have the mid-ir counterparts in our observations (Figure 1).  These two
 sources are oriented on an approximate NS line between two 1.2 mm peaks
 that fall on an EW line.  There appears to be
 a partial fan of H$_2$ emission surrounding IRS 3 whose apex would be to the north
 near IRS 1.  \citet{car99} had found 3.6 cm continuum at IRS 3; the data of \citet{sri02} show
 the 3.6 cm emission to be extended NE and W (see \citet{pal06} Fig. 1).  The large scale CO outflow
 has its axis oriented NS,  centered between IRS 1 and IRS 3, with red and blue lobes
 that are extended EW. 
 
  \citet{pal06} have observed this field on the SMA. They found
 a weak 1.3 mm dust peak and a CO(2-1) peak at IRS 1, with a compact EW bipolar CO
 outflow centered there.  They argued that IRS 1 is apparently a low/intermediate mass YSO. 
  They also suggested that the  redshifted CO lobe of \citet{beu02b} covering IRS 1 has a
  different underlying spatial scale  that the blueshifted CO lobe covering IRS 3, so that the
  two may not be directly related.
 Our mid-ir spectrum and 24.8 $\mu$m photometry for IRS 1 suggest an embedded YSO.
 Our model gives A$_V$=46 and a luminosity $L_e=1400$ L$_\sun$, consistent with a B3 ZAMS
 star that is too cool to create an HII region \citep{pan73} that we could detect.  
 The lack of an HII region is confirmed 
 by a lack of [Ne II] emission in addition to a lack of detected 3.6 cm emission.  This 
object is most likely an intermediate mass YSO that is
 a significant contributer to the IRAS-based luminosity of IRAS 20343.

The situation at IRS 3 seems ambiguous.  CS (2-1) observations of 20343 with a $27\arcsec$
beam gave a linewidth of 2.6 km s$^{-1}$ centered on 11.4  km s$^{-1}$ \citep{beu02a}. 
With the SMA, \citet{pal06} detected small amounts of 1.3 mm dust emission east and 
northwest of IRS 3, 
and low velocity CO (2-1) emission north, east, and west of IRS 3 with a 3.3 km s$^{-1}$ width.  
 The large-scale outflow red and blue shifts are separated by 
about 5.5 km s$^{-1} $ centered on 11.5 km s$^{-1} $
 \citep{beu02b}, but the red lobe does not appear to extend
to IRS 3.  In contrast, the [Ne II] observed FWHM is 8 km s$^{-1}$, indicating
an intrinsic width of 4-6 km s$^{-1}$, depending on the line shape, wider than the molecular
lines.  It is centered on $15.7\pm 1$ km s$^{-1} $, close to the velocity of the redshifted 
low velocity (13-15 km s$^{-1}$) molecular outflow
that is associated with IRS 1 only, rather than the blueshifted velocity ranges of
 either IRS 1 and IRS 3 \citep{pal06}. In
 our analysis, it is assumed that the centers of the 3.6 cm emission and the 
 [Ne II] emission are coincident.

 \citet{pal06} suggested that the large scale massive blue shifted CO lobe is associated
 with the fan of H$_2$ that is traced by lines of clumpy emission $\sim 10\arcsec$ 
 east and west of IRS 3 \citep{kum02}.  They argued that the emission around
 IRS 3 could be powered by a either
 a B2 star with
 an UC HII region or a lower mass YSO with an ionized outflow.  In either case, 
a cavity would have been created whose walls emit the 1.2 mm
 continuum emission condensations and the low velocity CO(2-1) seen by the SMA
 to the east and west, and the fan of 
 2.12 $\mu$m H$_2$ emission \citep{pal06}.

  We consider the ionized outflow model first. \citet{pal06} showed that a stellar wind 
 assumed 
 to have a 200 km s$^{-1}$ velocity  \citep{pal07,bel01}  could create the cavity.  There
 are  NE and W extensions of the 3.6
 cm emission shown in Fig. 1 of \citet{pal06} that would be consistent with jets  directed toward the cavity walls. The 24.8
  $\mu$m image (Fig. 1b) is extended parallel to the 
  3.6 cm extended emission to the NE, as would be expected if it traced a cavity wall, and
  there is very faint N band emission that appears to be associated with the H$_2$ emission shown
  in Fig. 1a. In contrast, the 
  GEMINI-TEXES 12.8 $\mu$m continuum map (Fig. 4) is extended NS rather than NE or W that would
 be expected if it traced the inner cavity wall around the 3.6 cm emission.  Compared to
 the NE-W 3.6 cm emission, the NS 12.8 $\mu$m continuum emission appears to be a possible 
 disk seen edge-on,
 but the [Ne II] emission is at the north end rather than at the center (Fig. 5).   If there were a 200 km s$^{-1}$ wind, we might expect to see a much wider [Ne II]
 line than the observed 8 km s$^{-1}$ either as part of the ionized wind or due
 to shock excitation.  The 15.7 km s$^{-1}$
 $V_{LSR}$ of the  [Ne II] is puzzling in this context in relation to the molecular cloud's
 velocity of 11.5 km s$^{-1}$ and lack of redshifted CO
 emission at IRS 3.
Finally, although the wind driven cavity hypothesis suggests
 the possibility of an intermediate mass protostar, the lack of reddening of IRS 3 in the 
 near-ir argues for a more evolved object  \citep{pal06}, and the lack of deep silicate absorption in the 
 mid-ir in our spectrum does as well.  
 
 As a likely alternative to a wind driven cavity \citet{pal06} suggested that cavity could have been
 cleared by radiation pressure of a B2 star.  The [Ne II]
 emission is consistent with the strength of the 3.6 cm emission of 1.8 mJy for the existence
 of an UC HII powered by a B2 star. The observed shift in $V_{LSR}$ of about 4 km s$^{-1}$
 relative to the molecular cloud is consistent with observed dispersion of stellar velocities in
 clusters, and models of UC HII regions within molecular
 clouds that sometimes  require relative velocities between the exciting star and its parent
 cloud \citep{zhu05}.
 The width of the [Ne II] line
 of 8 km sec$^{-1}$ suggests  small velocities of expansion, stellar wind driven motion, 
 or turbulence for an UC HII region.\footnote{
 The Doppler thermal width for Ne in an
8000 K HII region is 4.3 km s$^{-1}$.  Usually H recombination linewidths are quoted, for
which the Doppler thermal width is 19 km s$^{-1}$. UC HII regions typically show recombination
linewidths of 30-40 km sec$^{-1}$  \citep{hor07,zhu05}. }
While the TEXES [Ne II] map does not show the NE-W extension
 of the 3.6 mm continuum,  the cloudy model indicates
 that the H$^+$ zone should be more extended than the Ne$^+$ zone for a B2 star.    Our MIRSI mid-ir spectrum suggests an evolved envelope containing almost no hot dust.
The extended diffuse 8-13 $\mu$m N band emission with its marginally 
 resolved peak, and the marginally resolved 24.8 $\mu$m 
 peak with relatively large  flux are consistent with 
 partial destruction of the inner envelope that surrounded the star, even though
 the CS, CO and [Ne II] linewidths do not suggest current high velocity or 
 high luminosity outflows.  The different images in the N band,
  the TEXES 12.8 $\mu$m continuum (extended NS), and the 24.8 $\mu$m filter (extended
  NE) suggest remnant fragments  of the original disk and envelope.  The
 remnant star formation core has a mid-ir model luminosity of $L_e=850$ L$_{\sun}$,
 well below 3500 L$_\sun$ of a B2 star \citep{pan73}, but it may contribute more 
 than 850 L$_{\sun}$ to the 
 large-scale far-ir emission detected by IRAS.    This scenario
 that a B2 ZAMS powers an UC HII that creates the 3.6 cm and [Ne II] emission
  seems more likely than that of an intermediate mass protostar creating them 
 though a high velocity ionized wind .

 The high resolution mid-ir observations have identified the two most luminous sources
   in IRAS 20343+4129.  Together they can account for its IRAS luminosity. 
 

 \section{Summary and Conclusions}\label{conclusions}
 
 We have presented high resolution mid-ir observations made with
 MIRSI on the IRTF and TEXES on Gemini North of three  HMPO candidates  
 taken from a partial follow up survey of HMPO candidates originally studied 
at 1.2 mm and radio wavelengths by 
 \citet{sri02} and \citet{beu02a}.  They are typical for HMPO candidates observed 
 in the follow up survey being 
 compact at 1$\arcsec$ resolution,  having low resolution spectra with  strong, 
 moderate, or weak silicate absorption, and with  one  emitting
  the [Ne II] line.  
 
 A simple model of hot dust in emission,
 warm dust in emission, and cold dust in absorption was developed to fit our 8-13 
 $\mu$m low resolution spectra and our 24.8 $\mu$m photometric points.  Even an
 apparently flat 8-13 $\mu$m spectrum requires an absorption component if
 the underlying emission is assumed to be due to hot or warm silicate dust.  The
 temperatures ranged from $\sim$ 400-1000 K for the hot dust, and $\sim$ 100-200
 K for the warm dust.  Using \citet{dra03a} $R_V$=5.5 model dust properties and gas-to-dust ratio, 
 only small masses of gas and dust in the two emitting components
 are needed to fit the data.  The masses are less than about 1/10 solar mass (often
 much less) even though
 these are high or intermediate mass stars, and the mid-ir emission cannot be due
 the the bulk of the mass in massive accretion disks.   The mid-ir is likely to
 be emitted by the inner walls of outflow cavities and perhaps partly by the surfaces of
 accretion disks.
 On the other hand, high column densities, $10^{22} - 10^{23}$
 H nucleons cm$^{-2}$, are required for the cold absorption components.   These
 column densities are less than derived from 1.2 mm 11$\arcsec$ data using
 \citet{dra03a} dust, but the discrepancy may be resolved if the slope of the 
 absorption coefficient from far-ir to mm is flattened, as suggested by some 
 observations.   Our three component model is not meant to fit either near-ir or
 far-ir to mm ends of SEDs.  Nevertheless, the dust we are modeling in the
 hot and warm components appears to absorb the bulk of an HMPO's or intermediate mass
 YSO's  photospheric
 emission, so that the integrated flux of the two model components without application of the
 cold dust's extinction matches the luminosity as measured including the far-ir by IRAS.
The mid-ir measurements together with the model thus give a reasonable way to 
determine the luminosity for individual HMPOs.

The mid-ir emission of IRAS 18151-1208 together with weak 3.6 cm emission and 
other previous observations suggest that it is an early stage  pre-UC HII HMPO whose luminosity 
is that of a B0 star.

 TEXES high resolution spectra that cover emission lines from ionized gas
  can be used to determine the nature of the emission (jet or HII region) and 
  help
  determine the properties of the underlying star.   
 In the case of IRAS 20343+4129 IRS 1,  a lack of [NeII] emission, 
 a well defined compact CO outflow,
 a moderately strong silicate absorption feature,
 and  a dust
 model-based luminosity of 1400 L$_\sun$ imply that it is an
 intermediate mass YSO whose luminosity is that of  a B3 star.
 For IRAS 20343+4129 IRS 3 observed  [Ne II] emission and 
 3.6 cm free-free emission  are consistent with a cloudy model 
 indicating that the object is a B2 ZAMS star.  Its weak 
 silicate absorption and small mid-ir based luminosity suggest that it has already disrupted
 much of its natal envelope.  



\acknowledgments

H.B. acknowledges financial support by the Emmy-Noether-Program of the 
  Deutsche Forschungsgemeinschaft (DFG, grant BE2578).
We are grateful to R. T. Brooks of Colby College for work on improving the grism spectra reduction
process, and for  Dr. John W. Kuehne of Colby for system administration of the workstations used
for data acquisition, reduction, and modeling.  Miranda Harwarden-Ogata and Eric Tollestrup
provided tireless support from the IRTF for the remote observing program.
We thank M.J. Richter, T.K. Greathouse, M.A. Bitner, and D.T. Jaffe
for assistance with the TEXES observations.
This work is based in part on observations obtained at the Gemini Observatory,
which is operated by the Association of Universities for Research in
Astronomy, Inc., under a cooperative agreement with the NSF on behalf
of the Gemini partnership: the National Science Foundation (United States)
the Particle Physics and Astronomy Research Council (United Kingdom),
the National Research Council (Canada), CONICYT (Chile), the Australian
Resarch Council (Australia), CNPq (Brazil), and CONICET (Argentina).
Observations with TEXES were supported by NSF grant AST-0607312.


{\it Facilities:} \facility{IRTF (MIRSI)}, \facility{Gemini North (TEXES)}.

\clearpage

\clearpage

\begin{deluxetable}{lllrrrr}
\tablecolumns{7}
\tabletypesize{\scriptsize}
\tablecaption{MIRSI-IRTF Image Data\label{tbl-1}}
\tablewidth{0pt}
\tablehead{
\colhead{Object} & \colhead{RA(2000)} & \colhead{DEC(2000)} &
\colhead{F$_\nu$(10.5 $\mu$m)\tablenotemark{1}}& \colhead{F$_\nu$(24.8 $\mu$m)\tablenotemark{1}}&
\colhead{FWHM(10.5 $\mu$m)\tablenotemark{2} }& \colhead{FWHM(24.8 $\mu$m)\tablenotemark{2}}
}
\startdata
IRAS 18151-1208 & 18:17:58.1\tablenotemark{3}\tablenotemark{4}& -12:07:25.6\tablenotemark{3}
 \tablenotemark{4}& 11.2 & 101.4    &   1$\farcs$01 &  1$\farcs$91   \\
 IRAS 20343+4129 IRS1&20:36:7.6\tablenotemark{5}\tablenotemark{6}&41:40:08.0\tablenotemark{5}
 \tablenotemark{6}&8.1&15.6&0$\farcs$91&1$\farcs$88\\
  IRAS 20343+4129 IRS3&20:36:7.3\tablenotemark{5}\tablenotemark{7}&41:39:52.5\tablenotemark{5}
 \tablenotemark{7}&3.7&86.6&1$\farcs$24&2$\farcs$41\\
 $\gamma$ Aql&19:46:15.6&10:36:47.7&72.2&13.6&$0\farcs98$&$1\farcs71$\\
 PSF Average\tablenotemark{8}&\nodata&\nodata&\nodata&\nodata&0$\farcs$99&1$\farcs$83\\
 PSF Standard Deviation\tablenotemark{8}&\nodata&\nodata&\nodata&\nodata&0$\farcs088$&$0\farcs124$\\
\enddata
\tablenotetext{1} {Janskys.  Systematic uncertainty is $\pm10$\% for N band (10.5 $\mu$m),
and $\pm30$\% for 24.8 $\mu$m}
\tablenotetext{2}{ FWHM of Gaussian fit to image by IRAF task imexamine}
\tablenotetext{3}{3.6 cm source position \citep{car99}}
\tablenotetext{4}{IRTF indicated N position: 18:17:57.9 -12:07:27.0}
\tablenotetext{5}{K source position \citep{kum02}}
\tablenotetext{6}{IRTF indicated N position: 20:36:07.6 41:40:10.0}
\tablenotetext{7}{IRTF indicated N position: 20:36:07.3 41:39:54.0}
\tablenotetext{8}{Based on 6 observations of standard stars in N and 5 at 24.8 $\mu$m}
\end{deluxetable}

\clearpage
\begin{deluxetable}{lrrrl}
\tablecolumns{5}
\tabletypesize{\scriptsize}
\tablecaption{Source Diameters for Models\label{tbl-2}}
\tablewidth{0pt}
\tablehead{
\colhead{Object} & \colhead{Distance, pc \tablenotemark{1}}& \colhead{Hot Component, AU} &
\colhead{Warm Component, AU}& \colhead{Assumption for Calculation}
}

\startdata
IRAS 18151-1208 &3000&1280&2050&1$\sigma$ deviation for unresolved source\tablenotemark{2}\\
IRAS 18151-1208 &3000&2310&3670&3$\sigma$ upper limit for unresolved source\tablenotemark{2}\\
IRAS 20343+4129 IRS1&1400&600&960&1$\sigma$ deviation for unresolved source\\
IRAS 20343+4129 IRS1&1400&1080&1720&3$\sigma$ upper limit for unresolved source\\
IRAS 20343+4129 IRS3&1400&1120\tablenotemark{3}&2200&TEXES measurement (Hot) or \\
&&&&MIRSI Gaussian deconvolution (Warm)\tablenotemark{2}\\
IRAS 20343+4129 IRS3&1400&1590&2930&3$\sigma$ upper limit for resolved source\\
\enddata
\tablenotetext{1}{ \citet{sri02}}
\tablenotetext{2}{See text for details}
\tablenotetext{3}{Circular equivalent to TEXES-Gemini ellipse}
\end{deluxetable}

\clearpage

\clearpage

\begin{deluxetable}{lrrrrrrrrrrrrrrrr}
\tablecolumns{17}
\tabletypesize{\scriptsize}
\rotate
\tablecaption{Source Model Parameters\label{tbl-3}}
\tablewidth{0pt}
\tablehead{
\colhead{Object} &
 \colhead{$D_h$\tablenotemark{1} }& 
 \colhead{$T_{h}$} &
 \colhead{$\tau_{h}$}&
\colhead{log$N_{H,h}$ }&
\colhead{  log${  { M_{h} }\over{ M_{\sun}  }  }$ }&
 \colhead{$D_w$\tablenotemark{2} } &
  \colhead{ $T_{w}$ } &
  \colhead{$\tau_{w}$}&
\colhead{log$N_{H,w}$ }&
\colhead{  log${ { M_{w} }\over{ M_{\sun}  }   } $ }&
\colhead{ $\tau_{a}$\tablenotemark{3} }&
\colhead{ log$N_{H,a}$ }&
\colhead{ $A_{V,a} $}&
\colhead{log$L_{o}\over{ L_{\sun} } $\tablenotemark{4} }&
\colhead{log$L_{e}\over{ L_{\sun} }$\tablenotemark{5} }&
\colhead{$ \chi^{2}_{m} $}\\
\colhead{}&
\colhead{AU}&
\colhead{K}&
\colhead{}&
\colhead{}&
\colhead{}&
\colhead{ AU}&
\colhead{K}&
}

\startdata
 18151&1280.&455.&0.0545&21.05&-3.45&2050.&136.&4.20&22.93&-1.15&4.90&23.00&72.1&
 3.63&4.35&0.0045\\
 18151&1280.&1000.&0.0053&20.03&-4.46&2050.&171.&1.54&22.49&-1.60&4.85&23.00&71.3&
 3.70&4.71&0.0046\\
 20343 IRS1&600.&606.&0.0055&20.05&-5.10&960.&182.&0.12&21.38&-3.36&3.09&22.80&45.5
 &2.53&3.14&0.0005\\
 20343 IRS3&1120.&420.&0.0003&18.81&-5.80&2200.&110.&0.85&22.24&-1.78&0.79&22.21&11.6
 &2.85&2.93&0.0141\\

 \cutinhead{Upper Values}
 
  18151 &1280.&1000.&0.0691\tablenotemark{6}&21.15&-3.34&2050.&146.&
  7.90\tablenotemark{7}
  &23.21&-0.88&5.33&23.04&78.3&3.73&4.71&NA\tablenotemark{8}\\
 
  20343 IRS1&600.&666.&0.0072&20.17&-4.98&960.&201.&0.17&21.53&-3.21&3.47&22.85&51.0
   &2.58& 3.27&NA\\
 
 20343 IRS3&1220.&449.&0.0004&18.92&-5.69&2200.&112.&1.09&22.35&-1.67&1.14&22.37&16.8&2.95
 &3.02&NA\\ 
 
\cutinhead{Lower Values}

   18151 &1280.&422.&0.0036\tablenotemark{9}&19.87&-4.62&2050.&123.&0.94\tablenotemark{10}&22.65&-1.44&4.66&22.98&
   68.5&3.52&4.26&NA\tablenotemark{8}\\
   
    20343 IRS1&600.&533.&0.0037&19.88&-5.27&960.&159.&0.74&21.17&-3.57&2.80&
    22.75&41.3&2.40&3.01&NA\\
    
 20343 IRS3&1120.&380.&0.0002&18.62&-5.99&2200.&105.&0.58&22.07&-1.95&0.48&21.99&7.1
 &2.70&2.79&NA\\

\enddata
\tablenotetext{1}{\ Assumed diameter of hot component.}
\tablenotetext{2}{\ Assumed diameter of warm component.}
\tablenotetext{3}{\ Optical depth of cold extinction component.}
\tablenotetext{4}{\  Luminosity "observed" due to hot and warm dust components with 
extinction of cold component applied.}
\tablenotetext{5}{\  Luminosity emitted by hot and warm dust components without 
application of extinction of cold component.}
\tablenotetext{6}{\  Max  $\tau_h$ for $T_h=455$K. }
\tablenotetext{7}{\  Max $\tau_w$ for $T_w=136$K. The strongest effect is at 24.8 $\mu$m.}
\tablenotetext{8}{\  Values are derived from multiple models for which the parameter's upper
value shown gives
 $\chi^{2}_{m,i}\simeq0.09$ for the wavelength range most sensitive to the parameter.}
 \tablenotetext{9}{\  Min  $\tau_h$ for $T_h=1000$K.}
  \tablenotetext{10}{\  Min  $\tau_h$ for $T_w=171$K.}

\end{deluxetable}

\clearpage



\begin{figure}
\epsscale{.80}
\plotone{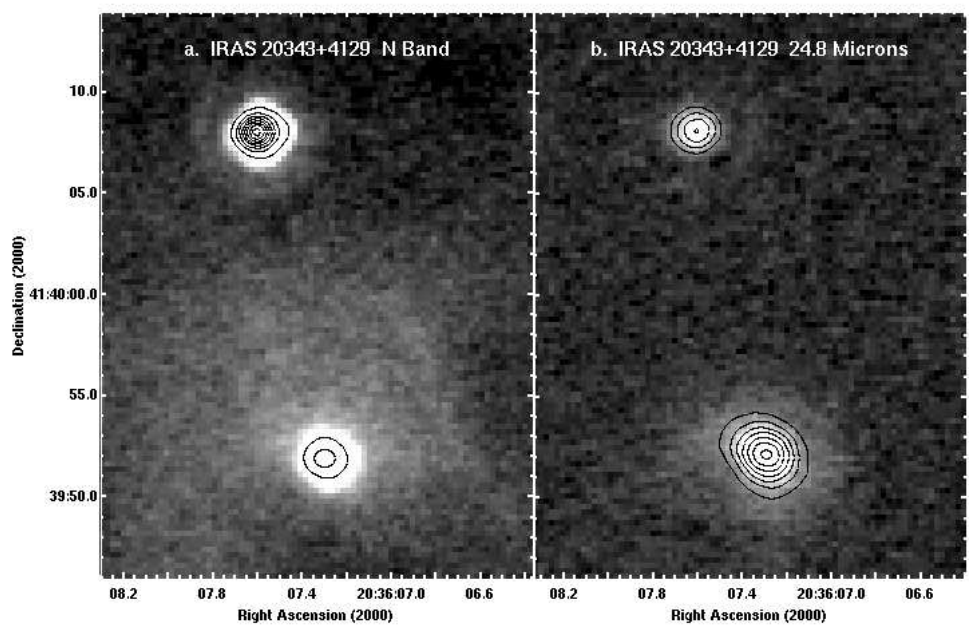}
\caption{IRAS 20343+4129 in N band centered at 10.5 $\mu$m ({\bf a})  and at 24.8 $\mu$m ({\bf b}).  IRS 1 is in the NE, and IRS 3 in the SW.  The grayscale is adjusted to show extended emission.  Eight equally spaced contours indicate the relative peak strength and extension for the central
regions of the sources.  Flux 
densities and FWHM are given in Table 1.  
   The true source positions are most likely to be at the
K band positions of IRS1: 20$^h$36$^m$7.6$^s$,
 41$\degr40\arcmin8\farcs0$ (J2000)
IRS3: 20$^h$36$^m$7.3$^s$,
 41$\degr39\arcmin52\farcs5$ (J2000)\citep{kum02}, and the coordinates shown have 
 been adjusted to fit them.  The 3.6 cm source of \citet{car99}
 is within an arcsecond of IRS 3.\label{fig1}}

\end{figure}

\clearpage
\begin{figure}
\epsscale{.80}
\plotone{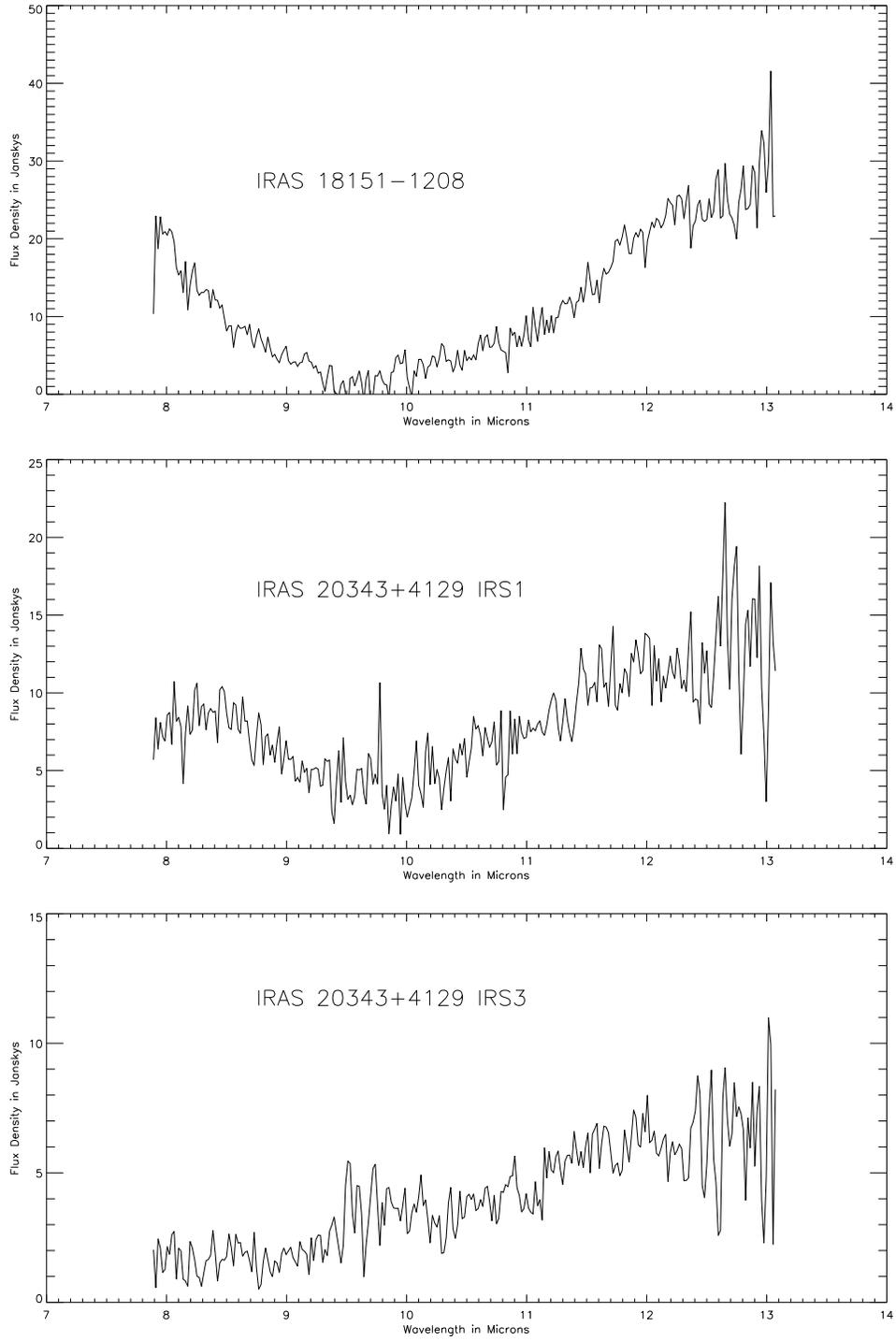}
\caption{Grism spectra of the three sources.\label{fig2}}
\end{figure}
\clearpage

\clearpage
\begin{figure}
\epsscale{.80}
\plotone{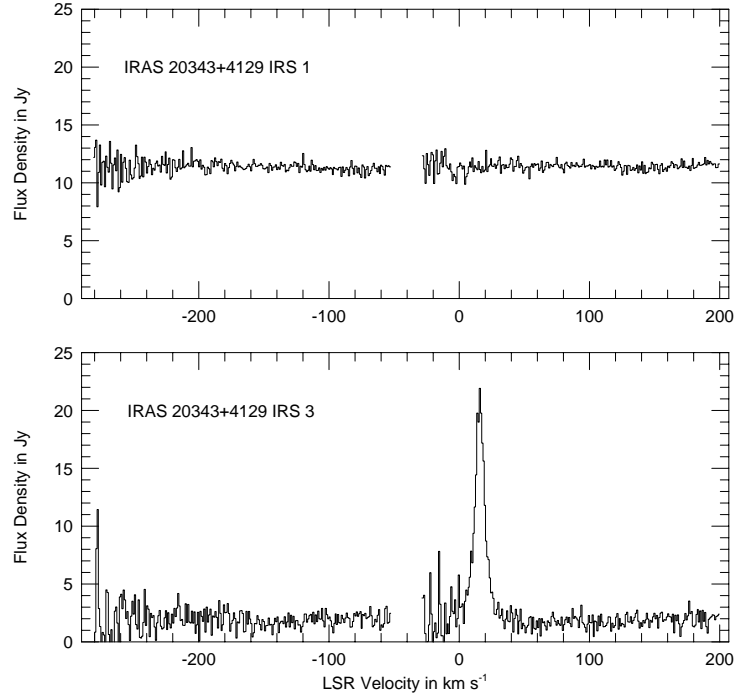}
\caption{TEXES spectra of IRAS 20343+4129 IRS 1 and IRS 3 for [Ne II] plotted
in velocity with respect to the local standard of rest.
The [Ne II] line in IRS 3 has its center at $V_{LSR} =15.7\pm1$  km s$^{-1}$  
($\lambda = 12.814\  \mu$m or $\nu = 780.383$ cm$^{-1}$ after correcting 
for the Earth's motion relative to the LSR). 
The gap in each spectrum at $V_{LSR} \sim-40$  km s$^{-1}$ is between grating orders. \label{fig3}}
\end{figure}
\clearpage

\clearpage
\begin{figure}
\epsscale{.80}
\plotone{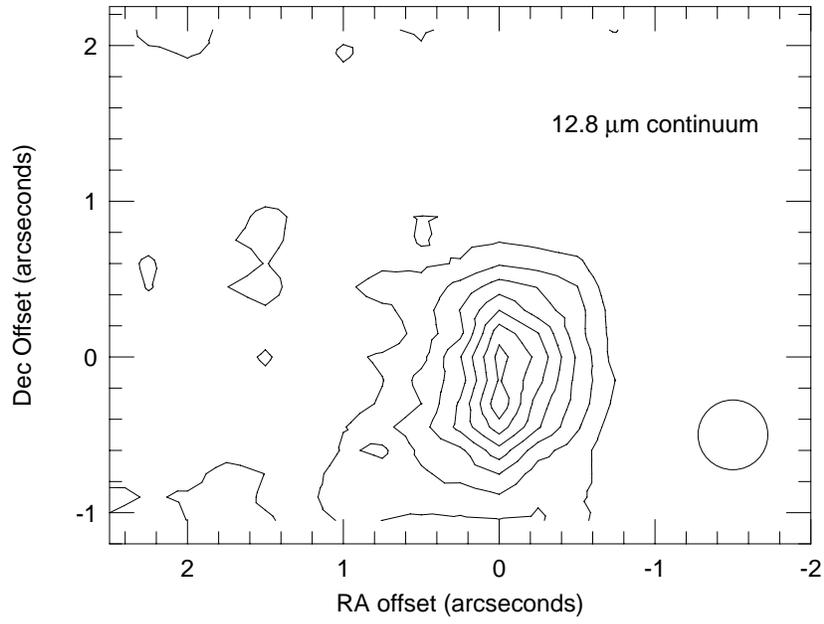}
\caption{ TEXES scan-mode 12.8 $\mu$m continuum map of IRAS 20343+4129
IRS 3.  The contours are at $1.67, 3.34, ...11.7 \times 10^{10} $Jy sr$^{-1}$
The beamsize of $0\farcs5$ is shown in the lower right.\label{fig4}}
\end{figure}
\clearpage

\clearpage
\begin{figure}
\epsscale{.80}
\plotone{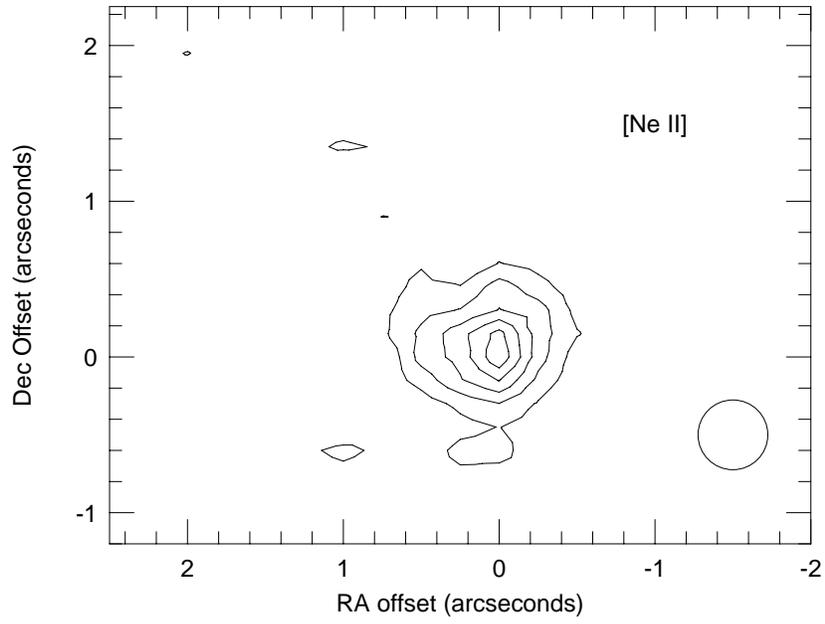}
\caption{TEXES scan-mode [Ne II] (12.8 $\mu$m) line map of IRAS
20343+4129 IRS 3.  
The contours are at $2.0, 4.0, ...10.0 \times 10^{-6} $ W  m$^{-2}$ sr$^{-1}$.
The beamsize of  $0\farcs5$ is shown in the lower right.\label{fig5}}
\end{figure}
\clearpage

\clearpage
\begin{figure}
\epsscale{.80}
\plotone{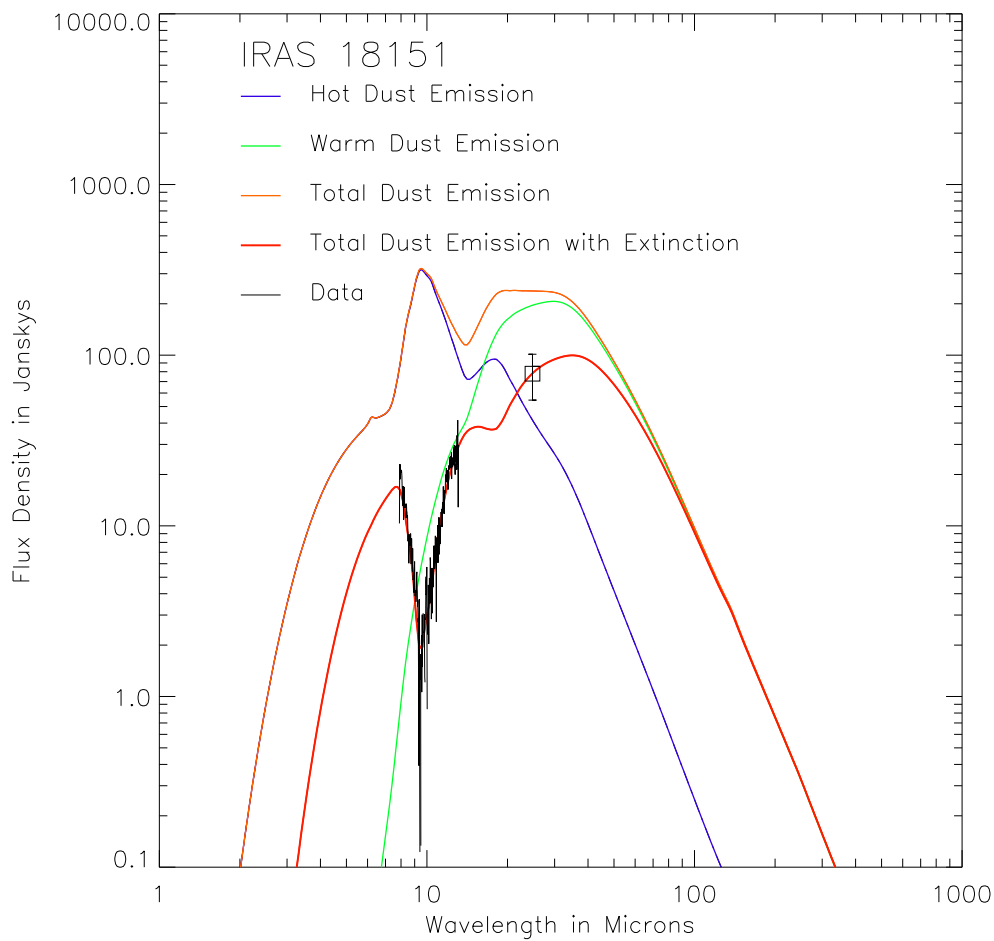}
\caption{Model individual and summed hot and warm components SEDs before extinction, and with
extinction for IRAS 18151-1208 compared to the MIRSI-IRTF data.  This model has 
a hot component temperature $T_h=455$ K.
Other parameters of the model are given in Table 3.  Emission from the hot and warm components 
before extinction
are shown in blue and green, respectively, and their sum, the total dust emission inside the outer cold 
cloud,
 is shown in orange.  The model's observed emission after extinction by the overlying cold cloud
 is shown in red, and the data are shown in black.\label{fig6}}
\end{figure}
\clearpage

\clearpage
\begin{figure}
\epsscale{.80}
\plotone{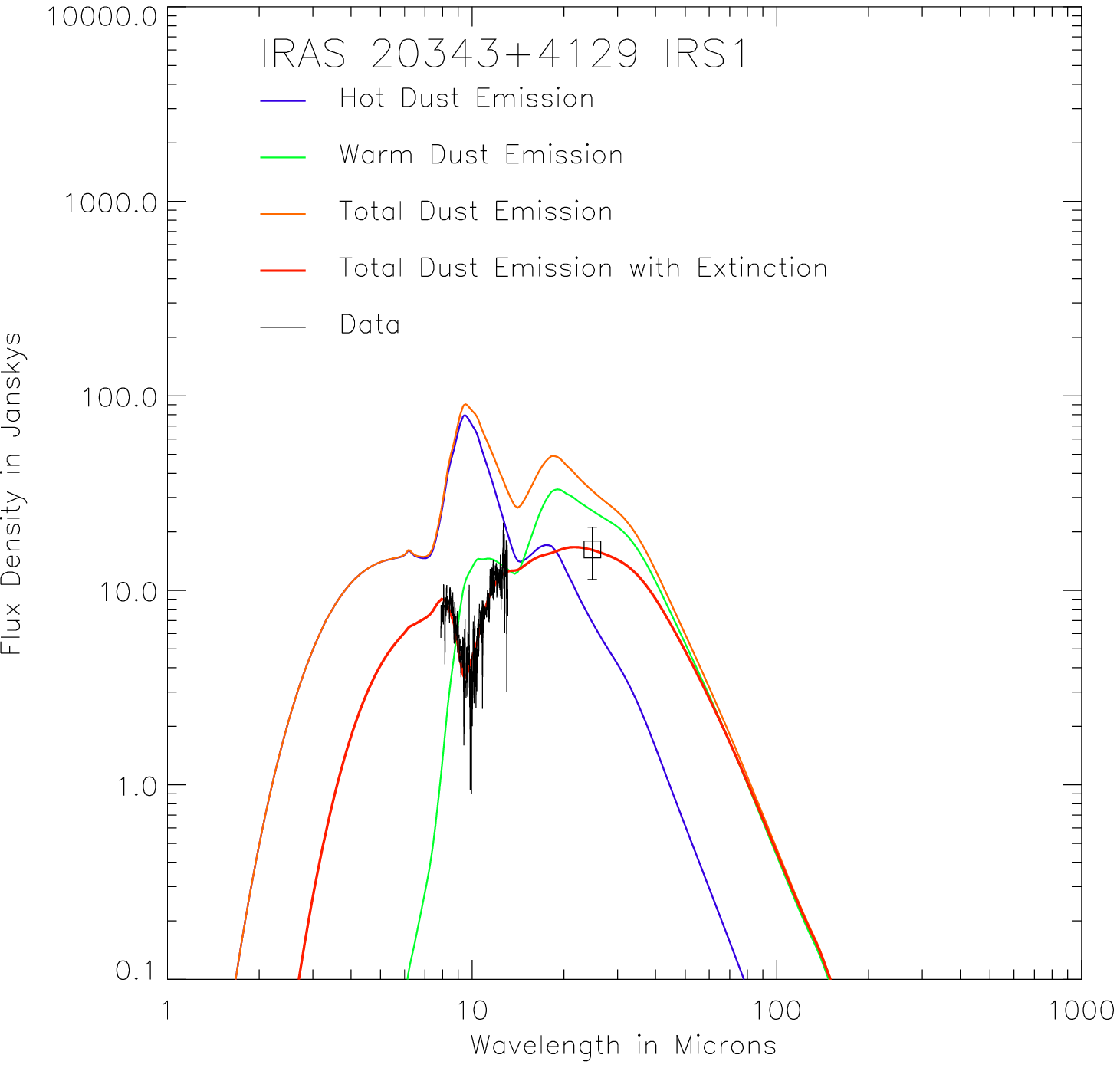}
\caption{Model individual and summed hot and warm components SEDs before extinction, and with
extinction for IRAS 20343+4129 IRS 1 compared to the MIRSI-IRTF data, as in Fig. 6.  Parameters of the model are given in Table 3. \label{fig7}}
\end{figure}
\clearpage

\clearpage
\begin{figure}
\epsscale{.80}
\plotone{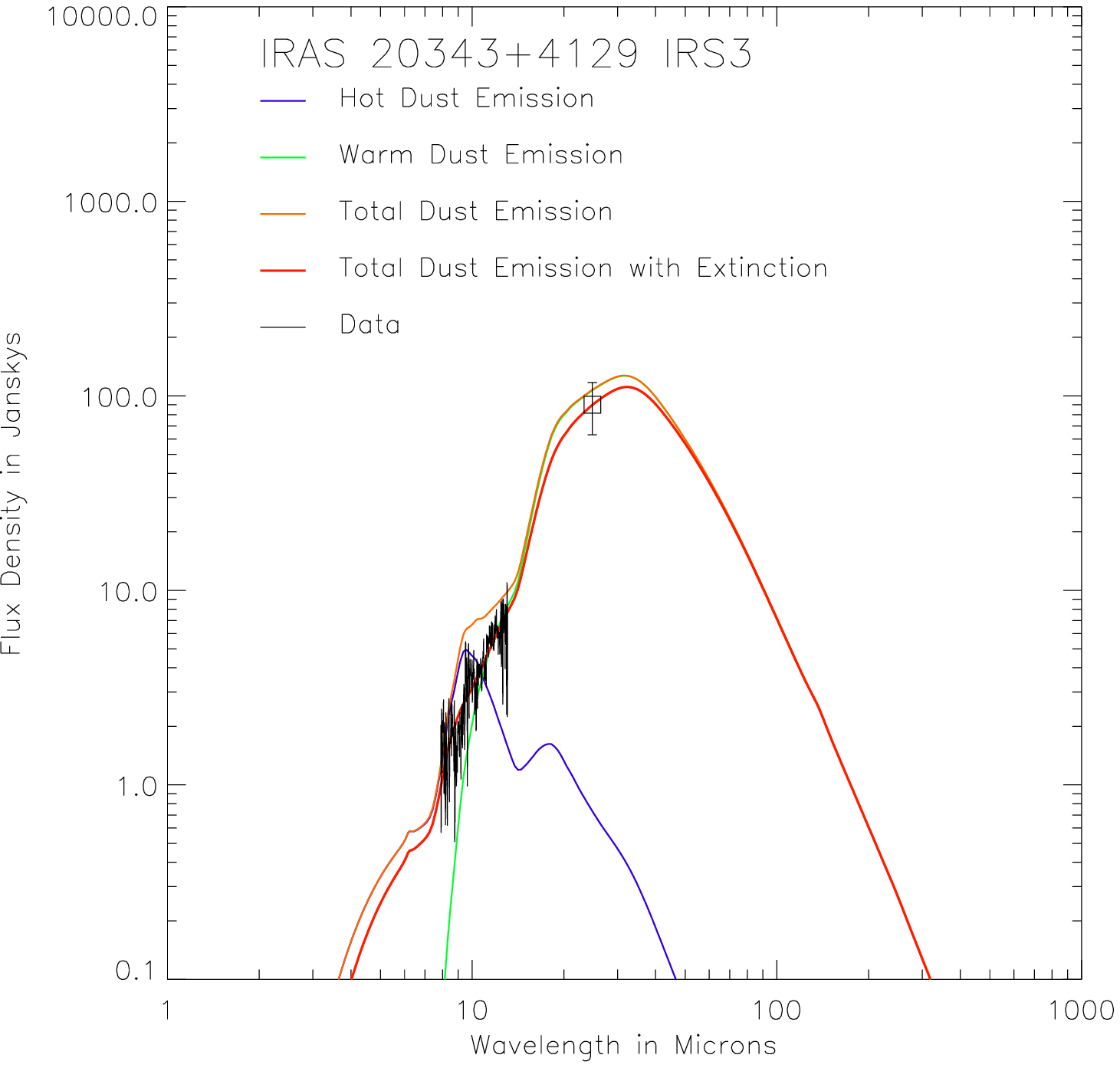}
\caption{Model individual and summed hot and warm components SEDs before extinction, and with
extinction for IRAS 20343+4129 IRS 3 compared to the MIRSI-IRTF data, as in Fig. 6.  Parameters of the model are given in Table 3. \label{fig8}}
\end{figure}
\clearpage

\end{document}